\documentclass[aip]{revtex4-1}
\usepackage{amsmath}
\usepackage{amssymb}
\usepackage{amsfonts}
\usepackage{latexsym}
\usepackage{color}
\usepackage{graphicx}
\newcommand{\stampa}[3]{\put(#1,#2){\linethickness{0pt}\makebox(0,0){#3}}}
\newcommand{\rettangolo}[6]{{\put(#1,#2){\line(1,0){#5}}}
                            {\put(#3,#2){\line(0,-1){#6}}}
                            {\put(#3,#4){\line(-1,0){#5}}}
                            {\put(#1,#4){\line(0,1){#6}}}
                           }

\newcommand{\ip}[2]{\left\langle {#1},{#2}\right\rangle}
\newcommand{\be}{\begin{equation}}
\newcommand{\en}{\end{equation}}
\newcommand{\bea}{\begin{eqnarray}}
\newcommand{\ena}{\end{eqnarray}}
\newcommand{\beano}{\begin{eqnarray*}}
\newcommand{\enano}{\end{eqnarray*}}
\newcommand{\bee}{\begin{enumerate}}
\newcommand{\ene}{\end{enumerate}}

\newcommand{\ST}{\mathcal{S}}

\newcommand{\Hil}{\mathcal{H}}

\newcommand{\Id}{1\!\!1}

\newcommand{\A}{\mathcal{A}}

\begin{document}
\title{Dynamics of closed ecosystems described by operators}

\author{Fabio Bagarello\footnote{Corresponding author. Tel.: +39-091-23897222, Fax: +39-091-427258.}}
\email{fabio.bagarello@unipa.it}
\affiliation{Dipartimento di Energia, Ingegneria dell'Informazione e Modelli Matematici, Facolt\`a di Ingegneria, Universit\`a di Palermo,
Viale delle Scienze, I--90128  Palermo, Italy}
\author{Francesco Oliveri}
\email{foliveri@unime.it}
\affiliation{Dipartimento di Matematica e Informatica, Universit\`a di Messina, Viale F. Stagno d'Alcontres 31, I--98166 Messina, Italy}
\email{foliveri@unime.it}

\date{\today}
\begin{abstract}
\noindent We adopt the so--called \emph{occupation number representation}, originally used in quantum mechanics and recently adopted in the
description of several classical systems, in the analysis of the dynamics of some models of closed ecosystems. In particular, we discuss two
linear models, for which the solution can be found analytically, and a nonlinear system, for which we produce numerical results. We also discuss how a damping effect could be {\em effectively} implemented in the model.
\end{abstract}


\maketitle

\section{Introduction and preliminaries}
\label{sect1}

In the past years it has been shown how an operatorial approach can be used to analyze very different classical systems. The leading idea of
this approach, which was recently reviewed in the monograph \cite{bagbook}, is that {\em raising} and {\em lowering} operators can be used to
describe some dynamical aspects of several classical systems, and in particular those systems whose main relevant variables (the {\em
observables} of the system) change discontinuously. For instance, this is what happens in stock markets, where the traders exchange an integer
number of shares, or in a predator-prey system, in which an integer number of preys are killed. Remarkably, the same technique, with minor
changes, seems to be useful also when continuous variables are involved Ref.~\onlinecite{ff2}.

In some  previous papers we have shown that an interesting difference does exist depending on the fact that the (eigenvalues of the)
observables of the system can take very high values or not. In the first case we have adopted bosonic operators, whereas in the second
situation we have used fermionic variables. This is related to the fact that the eigenvalues of a bosonic number operator are $0,1,2,3,\ldots$,
while 0 and 1 are the only allowed eigenvalues of a fermionic number operator. In the latter case, the analytical and numerical treatment of
the system appears much simpler than in the first situation. The reason is simple: the Hilbert space of the theory where fermionic number
operators live is finite dimensional, see Appendix A; it can be large, depending on the system we want to describe, but it is surely finite. On
the other hand, when we are forced to use bosonic operators,  even a very simple system is properly described in an infinite-dimensional
Hilbert space. However, we have shown \cite{ff2} that some conserved quantities can be used to define an {\em effective} finite-dimensional
Hilbert space, which contains all the relevant information about the system. There are other reasons for using fermionic operators besides the
intrinsic simplification related to the finite dimensionality of the Hilbert space; for instance, in Ref.~\onlinecite{ff3}, where a model of
interactive and migrating populations has been proposed, we have adopted fermionic operators since these mimic quite well densities (or {\em
local} densities) of species. This will be our choice also in this paper, which is devoted to describe the dynamical behavior of some different
models of closed ecosystems.

In recent years, the investigation about the environmental impact of mankind on the entire EarthÕs biosphere and on individual ecosystems \cite{PPBSS} has been
continuously increasing, either from an experimental or theoretical point of view.
Accurate experiments on natural ecosystems are difficult or practically impossible so that in many situations  small systems
laboratory ecosystems (involving only a limited number of components and often based on unicellular organisms) have been studied and mathematically modeled.
Closed ecological systems  \cite{BGMP,DeAngelis} are of special interest. These  are ecosystems that do not rely on matter exchange with any part outside the system. They are often used to describe small artificial systems designed and controlled by humans, \emph{e.g.}, agricultural systems and activated sludge plants, or aquaria, or fish ponds \cite{SKV}.
Mathematical models of such systems, besides being useful in describing the real earth's ecosystems, may help us in making predictions of how the system may change under certain circumstances. For artificial systems, models may help to optimize their design too. Artificial
closed ecosystems can potentially serve as a life support system during space flights, in space stations or space habitats \cite{DeAngelis}.
In a closed ecological system, any waste products produced by one species must be used by at least one other species and converted into nutrients: to do this an energy supply  from outside the system is needed. Therefore, a closed ecological system must contain at least one autotrophic --- chemotrophic or phototrophic --- organism.
Small closed ecosystems may serve as useful models for the analysis
of ecosystem properties in general, due to their relatively simple trophic
structure and the high intensity of the biotic material and energy transformations.
The most widely used mathematical models are compartment models whose time evolution is governed by a system of ordinary differential equations.

The models we want to present have a common structure: they are all made by $N$ different {\em internal} compartments (the {\em levels}),
interacting with a certain number of {\em external} compartments playing the role of the {\em nutrients} needed to feed the organisms in level
1 (autotroph organisms), and the {\em garbage} produced by all the elements occupying the various levels. Part of the garbage turns into
nutrients after some time. The organisms of levels greatest than 1 (heterotroph organisms) are feeded by those of the immediately preceeding
level. Each system considered here is closed, meaning with this that the only dynamical degrees of freedom are those of the levels, the
garbage(s) and the nutrients: there is nothing else, and only these quantities can interact between them. The simplest model is the one where
one has only a level of heterotroph organisms. It is probably worth stressing that the models we are going to consider are a first simplified
version of what a realistic closed ecological system should be. However, we believe they are good starting points to check whether the approach
we are adopting could be of some utility even for more complicated and detailed models.

The paper is organized as follows. In Section \ref{sect2}, we introduce a simple linear model, describing $N$ different levels of the system,
interacting with two external compartments playing the role of the {\em nutrients} and the {\em garbage}. In Section \ref{sect3}, we consider another linear
model with two different garbages. These are intended to model the fact that part of the garbage turns into nutrients quite fast, while for another
part this change could be much slower. Section \ref{sect4} describes a nonlinear version of the same system, while, in Section \ref{sect5}, we introduce
phenomenologically a damping effect to model the fact that, after a sufficiently long time, if the ecosystem is supposed to be unable to recycle completely all the produced garbage, the densities of
the species are expected to decrease signicantly, and to approach to zero eventually. Our conclusions are contained in Section \ref{sect6}.  Appendix \ref{AppendixA} contains few fact on fermionic operators,
useful to keep the paper self-contained. In Appendix \ref{AppendixB} we discuss a simple model with phenomenological damping, which motivates what we have
done in Section \ref{sect5}.

\section{A linear model with a single garbage}
\label{sect2}

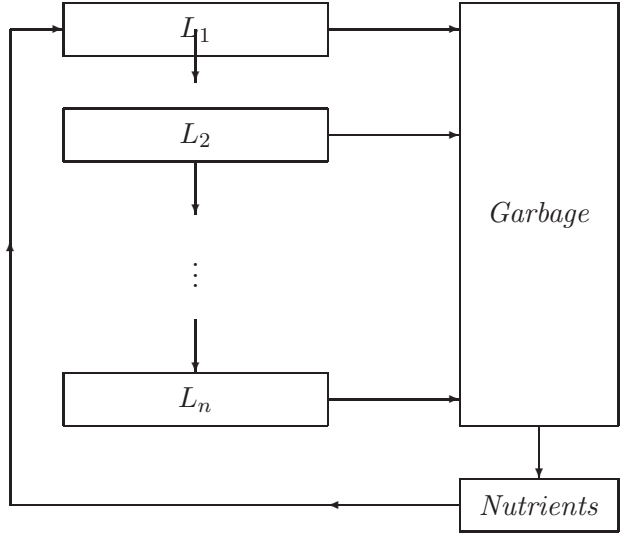
\begin{figure}
\begin{center}
\begin{picture}(400,210)(0,0)
\rettangolo{100}{205}{200}{185}{100}{20}
\stampa{150}{195}{$L_1$}
\put(150,195){\vector(0,-1){20}}
\put(200,195){\vector(1,0){50}}
\rettangolo{100}{165}{200}{145}{100}{20}
\stampa{150}{155}{$L_2$}
\put(150,145){\vector(0,-1){20}}
\put(200,155){\vector(1,0){50}}
\stampa{150}{105}{$\vdots$}
\put(150,85){\vector(0,-1){20}}
\rettangolo{100}{65}{200}{45}{100}{20}
\stampa{150}{55}{$L_n$}
\put(200,55){\vector(1,0){50}}
\rettangolo{250}{205}{310}{45}{60}{160}
\stampa{280}{125}{\emph{Garbage}}
\put(280,45){\vector(0,-1){20}}
\rettangolo{250}{25}{310}{5}{60}{20}
\stampa{280}{15}{\emph{Nutrients}}
\put(250,15){\vector(-1,0){50}}
\put(200,15){\line(-1,0){120}}
\put(80,15){\line(0,1){50}}
\put(80,65){\vector(0,1){50}}
\put(80,115){\line(0,1){80}}
\put(80,195){\vector(1,0){20}}
\end{picture}
\end{center}
\caption{\label{fig1}\footnotesize A schematic view to the single-garbage ecosystem.}
\end{figure}

The first closed ecosystem which we consider here is also the simplest one, made of $N$ levels of organisms,  one compartment for the nutrients
and a single compartment for the garbage, see Figure \ref{fig1} for a schematic view. Since we are interested in the densities of these
compartments, we adopt here (and in the next sections) fermionic operators, as we have successfully done in Ref.~\onlinecite{ff3}.

The dynamics (see Ref.~\onlinecite{bagbook}) is described by a hamiltonian operator containing the essential features of the system  we want to model.
Here we use the following hamiltonian:
 \be
\left\{
\begin{aligned}
&H=H_0+ H_I, \qquad \hbox{ with }  \\
&H_0 = \sum_{j=0}^{N+1}\,\omega_{j}\, a_j^\dagger\,a_j,\\
&H_I = \sum_{j=0}^N\,\lambda_j\left(a_j\,a_{N+1}^\dagger+a_{N+1}a_j^\dagger\right)+
 \sum_{j=0}^{N-1}\,\nu_j\left(a_j\,a_{j+1}^\dagger+a_{j+1}a_j^\dagger\right),
\end{aligned}
\right. \label{21}
\en
where \be \{a_j,a_k^\dagger\}=\delta_{j,k}\Id,\qquad a_j^2=0,\label{22} \en for all $j, k=0,1,\ldots,N+1$, and where
$\omega_j, \nu_j$ and $\lambda_j$ are real constants. The zero-th mode is related to the nutrients, the $(N+1)$-th mode to the garbage, while
all the remaining modes describe the organisms of the various trophic levels. The hamiltonian (\ref{21}) contains a free {\em standard} part,
$H_0$, whose parameters measure the inertia of the different compartments \cite{bagbook}: the higher the value of a certain $\omega_{j}$, the
higher the tendency of the density of the $j$-th degree of freedom to stay constant in time, even in presence of interaction. The next term,
$H_I$, which is  quadratic in the raising and lowering operators, describes the following effects: $\lambda_j\,a_j\,a_{N+1}^\dagger$ describes
an increasing of garbage and a simultaneous decreasing of the densities of the levels ($j=1,2,\ldots,N$): metabolic waste and death organisms
become garbage! For $j=0$, $H_I$ contains a similar contribution, $\lambda_j a_{N+1}a_j^\dagger$, describing the fact that the garbage is
recycled by decomposers and transformed into nutrients. Recall that, to make the hamiltonian self-adjoint, we are also forced to add the
adjoint contributions. The term $\nu_j\,a_j\,a_{j+1}^\dagger$ describes the fact that the nutrients are used by the organisms of level 1, and
that the organisms of level $j$ feed those of the level $j+1$ ($j=1,\ldots,N-1$). Again, the adjoint contribution $\nu_ja_{j+1}a_j^\dagger$
needs to be inserted in $H_I$.

The equations of motion, deduced by $\dot X=i[H,X]$, are therefore:
\be
\left\{
\begin{aligned}
&\dot a_0=i\left(-\omega_0a_0+\lambda_0a_{N+1}+\nu_0a_1\right),  \\
&\dot a_l=i\left(-\omega_la_l+\lambda_la_{N+1}+\nu_{l-1}a_{l-1}+\nu_la_{l+1}\right),\\
&\dot a_N=i\left(-\omega_Na_N+\lambda_Na_{N+1}+\nu_{N-1}a_{N-1}\right),\\
&\dot a_{N+1}=i\left(-\omega_{N+1}a_{N+1}+\sum_{j=0}^{N}\lambda_ja_{j}\right),
\end{aligned}
\right.
\label{23}
\en
where $l=1,2,\ldots,N-1$. Recall that $a_0$ and $a_{N+1}$ are not organisms but the nutrients and the garbage, respectively. It
is not surprising, therefore, that the related equations of motion differ from the other ones. Also, the equation for $a_N$ looks slightly
different from those for $a_l$, $l=1,2,\ldots,N-1$, since the $N-$th level has a single outgoing arrow, which goes to the garbage.

System (\ref{23}) can be rewritten as $\dot A=XA$, where
$$
A=\left(
    \begin{array}{c}
      a_0 \\
      a_1 \\
      \vdots \\
      \vdots \\
      a_N \\
      a_{N+1} \\
    \end{array}
  \right),\qquad X=i\,\left(
                     \begin{array}{cccccc}
                       -\omega_0 & \nu_0 & 0 & \cdots & \cdots & \lambda_0 \\
\nu_0 & -\omega_1 & \nu_1 & \cdots & \cdots & \lambda_1 \\
\cdots & \cdots & \cdots & \cdots & \cdots & \cdots \\
\cdots & \cdots & \cdots & \cdots & \cdots & \cdots \\
\cdots & \cdots & \cdots & \cdots & -\omega_N & \lambda_N \\
\lambda_0 & \lambda_1 & \lambda_2 & \cdots & \lambda_N & -\omega_{N+1} \\
\end{array}
\right),
$$
$X$ being a symmetric matrix. The solution is $A(t)=V(t)A(0)$, with $V(t)=\exp(Xt)$. Calling $V_{k,l}(t)$ the entries of the matrix $V(t)$, and $n_k(t)=\ip{\varphi_{\bf n}}{a_k^\dagger(t)a_k(t)\varphi_{\bf n}} $, where $\mathbf{n}=(n_0,n_1,\ldots,n_N,n_{N+1})$ are
the initial conditions (see Ref.~\onlinecite{bagbook}), we find that
\be
n_k(t)=\sum_{l=0}^{N+1}\left|V_{k,l}(t)\right|^2\,n_l.\label{24}
\en
These are the required densities of the
various compartments of the system, $k=0,1,2,\ldots,N+1$, with initial conditions fixed by the vector
$\varphi_{\mathbf{n}}$.

The explicit form of $n_k(t)$ is not very interesting for us, and will not be investigated further in this paper\footnote{In principle, it is
not hard to deduce its analytical expression, especially for small values of $N$.}, since the model we have considered is simply a first
approximation of the one we have in mind, which is in some sense more realistic since it models the possibility of having garbages of different
kind, in particular a {\em soft garbage}, which easily turns into nutrients, and a {\em hard garbage}, which also produces nutrients but only
after a much longer period. For instance, the soft garbage could come mainly from autotroph organisms, and hard garbage from heterotroph
organisms.

\section{A linear model with two garbages}
\label{sect3}

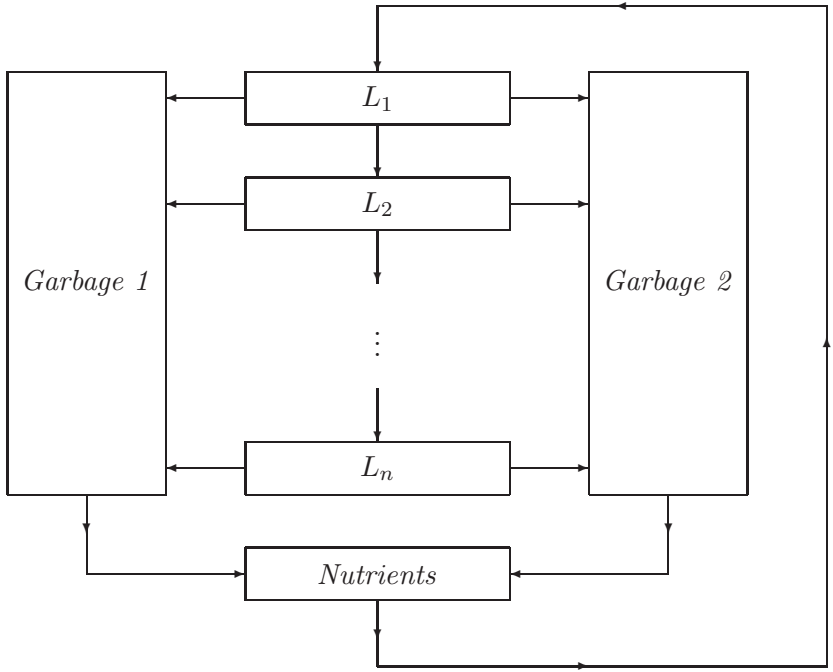
\begin{figure}
\begin{center}
\begin{picture}(400,250)(0,0)
\rettangolo{100}{225}{200}{205}{100}{20}
\stampa{150}{215}{$L_1$}
\put(150,205){\vector(0,-1){20}}
\put(200,215){\vector(1,0){30}}
\put(100,215){\vector(-1,0){30}}
\rettangolo{100}{185}{200}{165}{100}{20}
\stampa{150}{175}{$L_2$}
\put(150,165){\vector(0,-1){20}}
\put(200,175){\vector(1,0){30}}
\put(100,175){\vector(-1,0){30}}
\stampa{150}{125}{$\vdots$}
\put(150,105){\vector(0,-1){20}}
\rettangolo{100}{85}{200}{65}{100}{20}
\stampa{150}{75}{$L_n$}
\put(200,75){\vector(1,0){30}}
\put(100,75){\vector(-1,0){30}}
\rettangolo{10}{225}{70}{65}{60}{160}
\stampa{40}{145}{\emph{Garbage 1}}
\put(40,65){\vector(0,-1){15}}
\put(40,50){\line(0,-1){15}}
\put(40,35){\vector(1,0){60}}
\rettangolo{230}{225}{290}{65}{60}{160}
\stampa{260}{145}{\emph{Garbage 2}}
\put(260,65){\vector(0,-1){15}}
\put(260,50){\line(0,-1){15}}
\put(260,35){\vector(-1,0){60}}
\rettangolo{100}{45}{200}{25}{100}{20}
\stampa{150}{35}{\emph{Nutrients}}
\put(150,25){\vector(0,-1){15}}
\put(150,10){\line(0,-1){10}}
\put(150,0){\vector(1,0){80}}
\put(230,0){\line(1,0){90}}
\put(320,0){\vector(0,1){125}}
\put(320,125){\line(0,1){125}}
\put(320,250){\vector(-1,0){80}}
\put(240,250){\line(-1,0){90}}
\put(150,250){\vector(0,-1){25}}
\end{picture}
\end{center}
\caption{\label{fig2}\footnotesize A schematic view to the two-garbages ecosystem.}
\end{figure}

Let us now add a second reservoir to the system. The leading idea is that, considering different coupling constants between the garbages $G_1$
and $G_2$ with the nutrients $F$, we will be able to model the fact that part of the waste products and dead organisms is turned into nutrients
quickly (say, the autotroph detritus), while other parts (say, the heterotroph detritus) are converted in nutrients only after a longer time.
The presence of two compartments for the garbages imply that we need to add another degree  of freedom and, consequently, an extra fermionic
operator $a_{N+2}$. The other ingredients, as well as their meaning, are those of the previous model. The structure of the ecosystem is
depicted in Figure \ref{fig2}. The main difference with respect to the system described in Section \ref{sect2}, and with what shown in Figure
\ref{fig1}, is that two arrows now start from each level $L_j$, moving towards $G_1$ and $G_2$. Moreover, both $G_1$ and $G_2$ (with different
time scales) contribute to the nutrients.

\begin{figure}[h]
\begin{center}
\includegraphics[width=0.47\textwidth]{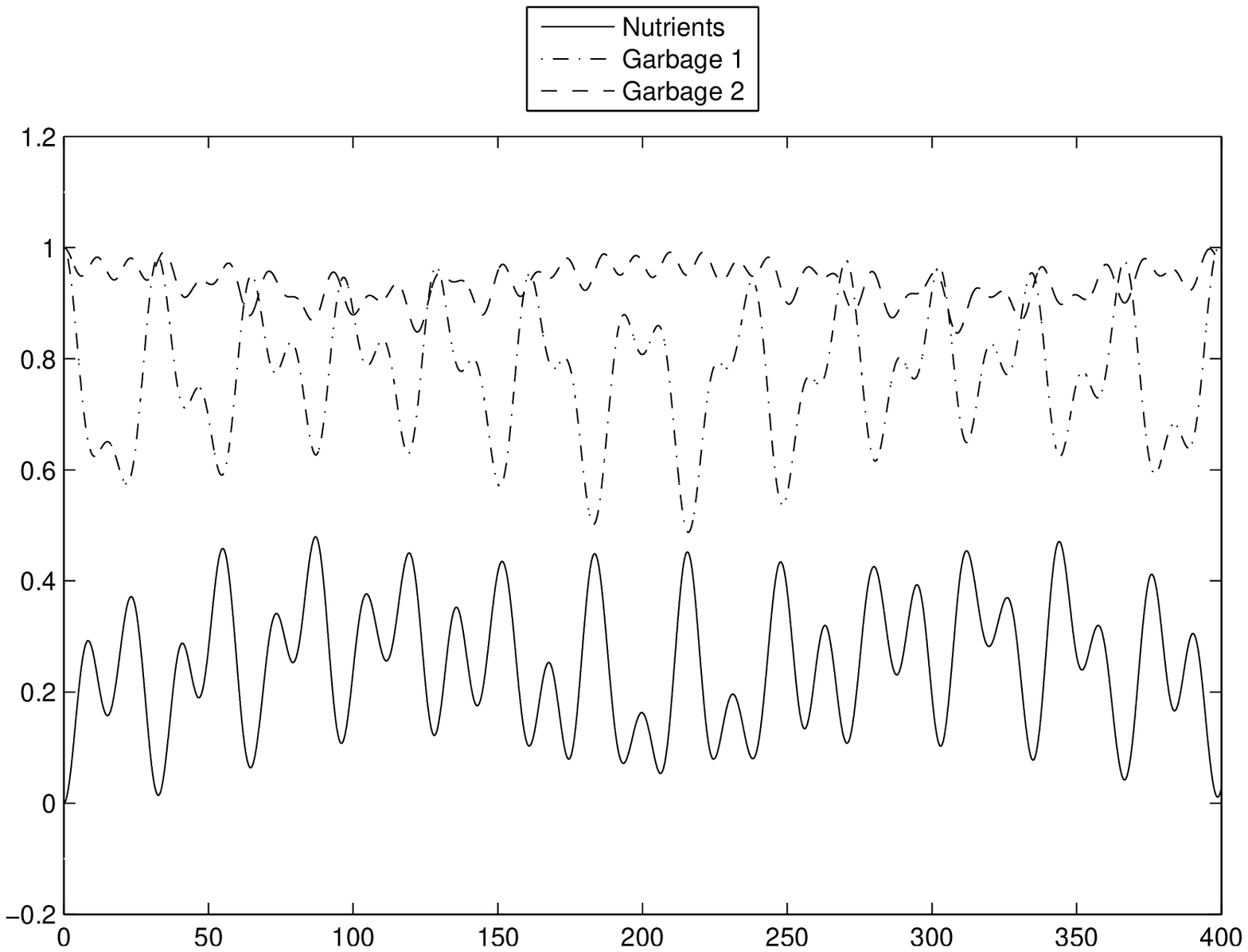}\hspace{8mm}
\includegraphics[width=0.47\textwidth]{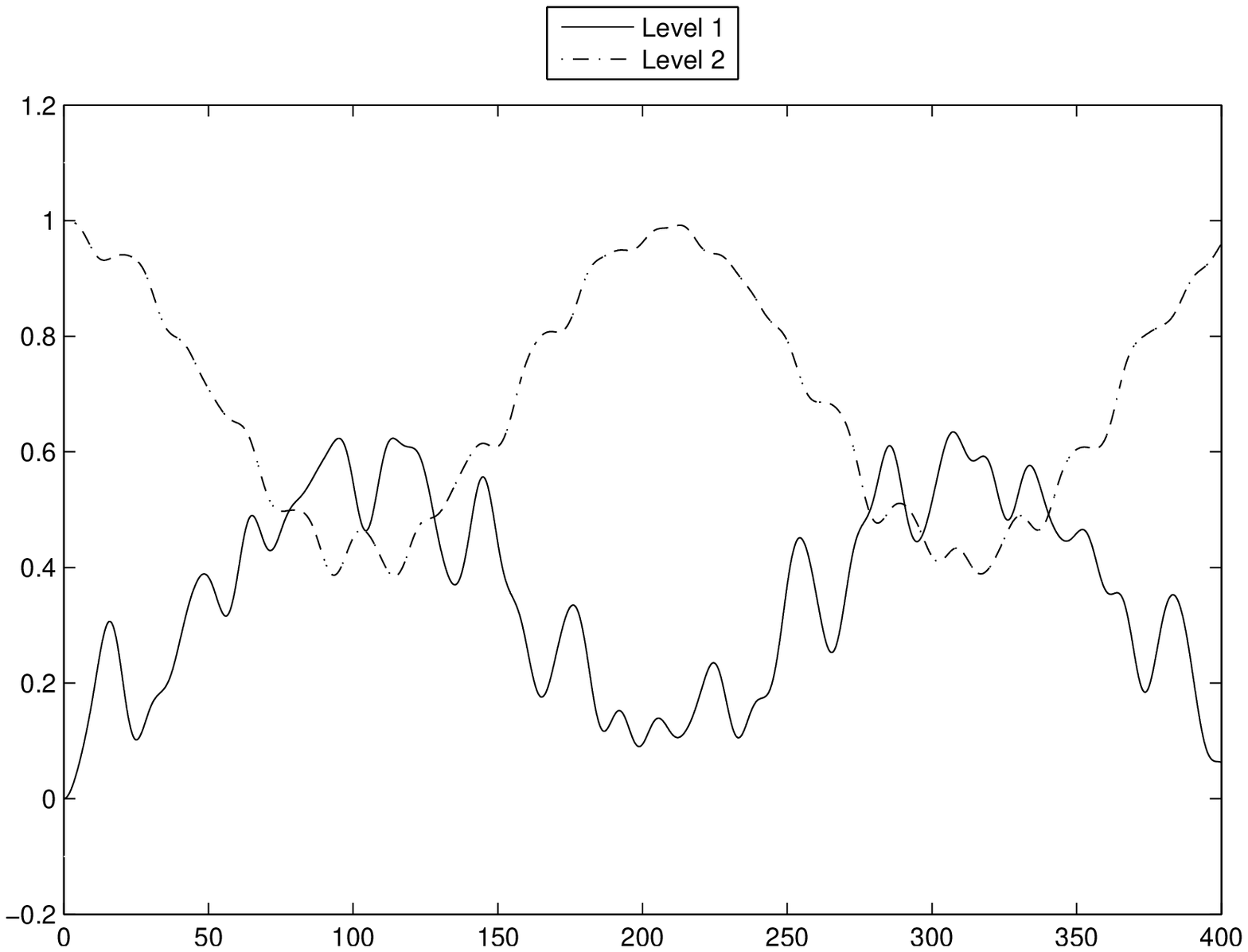}\hfill\\
\caption{\label{fig3a}\footnotesize Densities of the nutrients, $G_1$ and $G_2$, left, and of Levels 1 and 2, right. Initial conditions: nutrients and level 1 empty,
$G_1$, $G_2$ and level 2 completely filled.}
\end{center}
\end{figure}

The hamiltonian now looks like
\be
\left\{
\begin{aligned}
&H=H_0+ H_I, \qquad \hbox{ with }  \\
&H_0 = \sum_{j=0}^{N+2}\,\omega_{j}\, a_j^\dagger\,a_j,\\
&H_I = \sum_{j=0}^N\,\lambda_j^{(1)}\left(a_j\,a_{N+1}^\dagger+a_{N+1}a_j^\dagger\right)+
 \sum_{j=0}^N\,\lambda_j^{(2)}\left(a_j\,a_{N+2}^\dagger+a_{N+2}a_j^\dagger\right)  \\
&\quad+\sum_{j=0}^{N-1}\,\nu_j\left(a_j\,a_{j+1}^\dagger+a_{j+1}a_j^\dagger\right),
\end{aligned}
\right.
\label{31}
\en
where $\lambda_j^{(1)}$ describes the interaction between the organisms and $G_1$, while $\lambda_j^{(2)}$ is used to
fix the strength of the interaction between the organisms and $G_2$. The meaning of the various contributions are analogous to those in Section \ref{sect2}, and will not be repeated here. In particular, the last term in $H_I$ is identical to a contribution already appearing in (\ref{21}). The equations of motion extend those in (\ref{23}),
\be
\left\{
\begin{aligned}
&\dot a_0=i\left(-\omega_0a_0+\lambda_0^{(1)}a_{N+1}+\lambda_0^{(2)}a_{N+2}+\nu_0a_1\right),  \\
&\dot a_l=i\left(-\omega_la_l+\lambda_l^{(1)}a_{N+1}+\lambda_l^{(2)}a_{N+2}+\nu_{l-1}a_{l-1}+\nu_la_{l+1}\right),\\
&\dot a_N=i\left(-\omega_Na_N+\lambda_N^{(1)}a_{N+1}+\lambda_N^{(2)}a_{N+2}+\nu_{N-1}a_{N-1}\right),\\
&\dot a_{N+1}=i\left(-\omega_{N+1}a_{N+1}+\sum_{l=0}^N\lambda_l^{(1)}a_{l}\right),\\
&\dot a_{N+2}=i\left(-\omega_{N+2}a_{N+2}+\sum_{l=0}^N\lambda_l^{(2)}a_{l}\right),
\end{aligned}
\right. \label{32} \en $l=1,2,\ldots,N-1$, and can be solved in a similar way. Setting $N_{tot}:=\sum_{l=0}^{N+2}a_l^\dagger\,a_l$, we can
check that $[H,N_{tot}]=0$, so that $N_{tot}$ is a conserved quantity: what disappears from the levels appears in the garbages and in the
nutrients. To make the situation easy but not trivial, let us fix $N=2$: such a simplifying choice corresponds to identify levels 1 and 2 with
the autotroph and  heterotroph organisms, respectively. In Figure \ref{fig3a} we show how the densities of the various compartments of the
system change in time, where the parameters are set as follows: $\lambda_1^{(1)}=0.005$, $\lambda_2^{(1)}=0.009$, $\lambda_1^{(2)}=0.05$,
$\lambda_2^{(2)}=0.09$, $\nu_0=0.1$, $\nu_1=0.01$, $\nu_2=0.1$, $\omega_0=0.05$, $\omega_1=0.1$, $\omega_2=0.2$, $\omega_3=0.3$ and
$\omega_4=0.45$. This particular choice is motivated by the following reasons: since $G_2$ is the hard garbage, while  $G_1$ is the soft one,
it is clear that the inertia of $G_2$, measured by $\omega_4$, must be larger than that of $G_1$, which is measured by $\omega_3$. For this
reason we have taken $\omega_4>\omega_3$\footnote{We will come back on this aspect later on.}. Moreover, since the nutrients should be easily
used by the organisms, $\omega_0$ is taken to be very small: almost no inertia. Levels 1 and 2 are distinguished by assuming a larger inertia
for level 2 with respect to that of level 1: $\omega_1<\omega_2$. The first level interacts with $G_1$ and $G_2$ at a rate less than that of
the second one: for this reason we are taking $\lambda_1<\lambda_2$. $\nu_1$ is very small, compared with $\nu_0$, because organisms of level 1
use the nutrients to increase their density with time scales smaller than those needed by the organisms of level 2 which are feeded by the
organisms of level 1. In Figure \ref{fig3a} we are assuming that the nutrients and level 1 are empty (say, very low densities) at $t=0$, while
the two garbages and level 2 are completely filled (say, very high densities).

\begin{figure}[h]
\begin{center}
\includegraphics[width=0.47\textwidth]{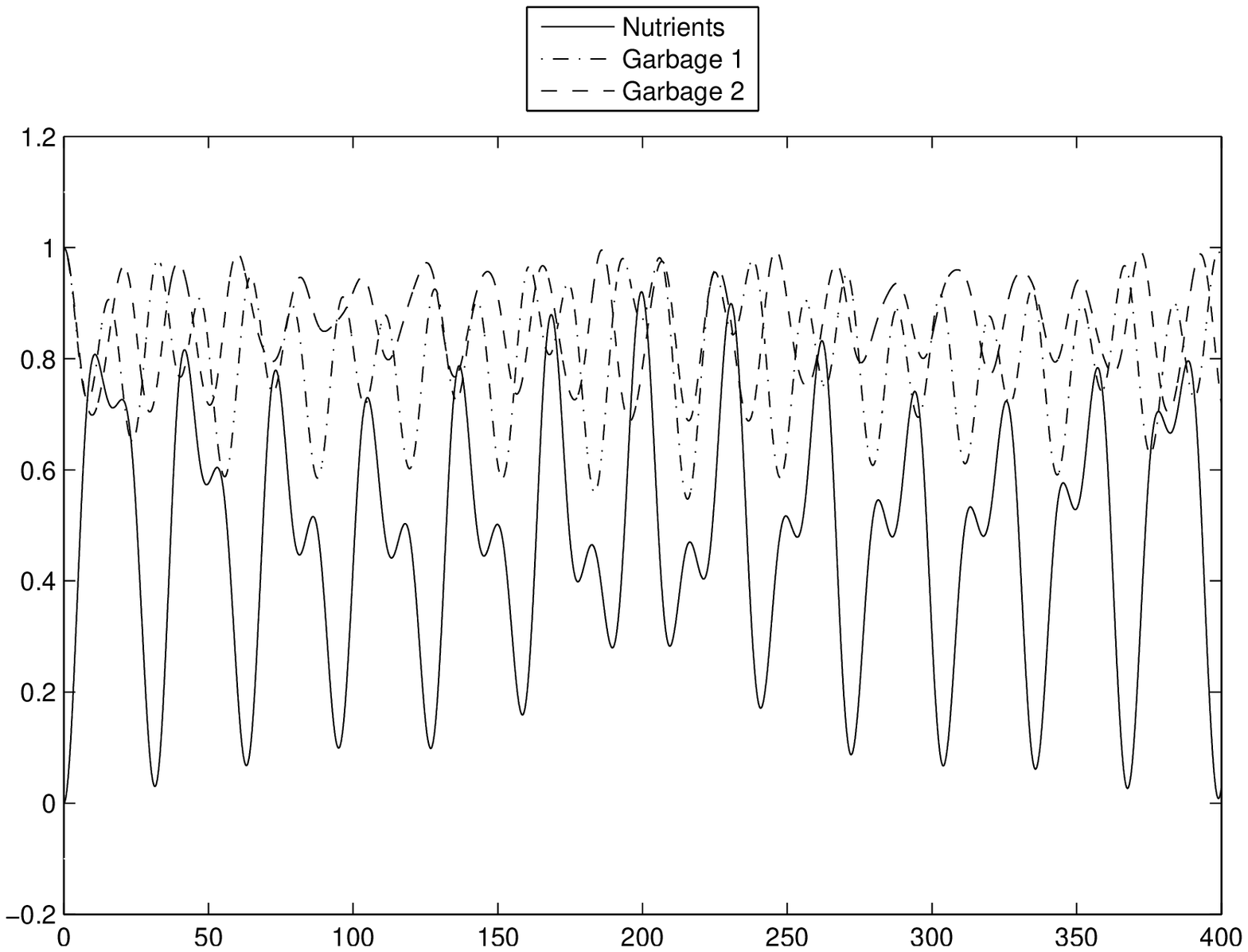}\hspace{8mm}
\includegraphics[width=0.47\textwidth] {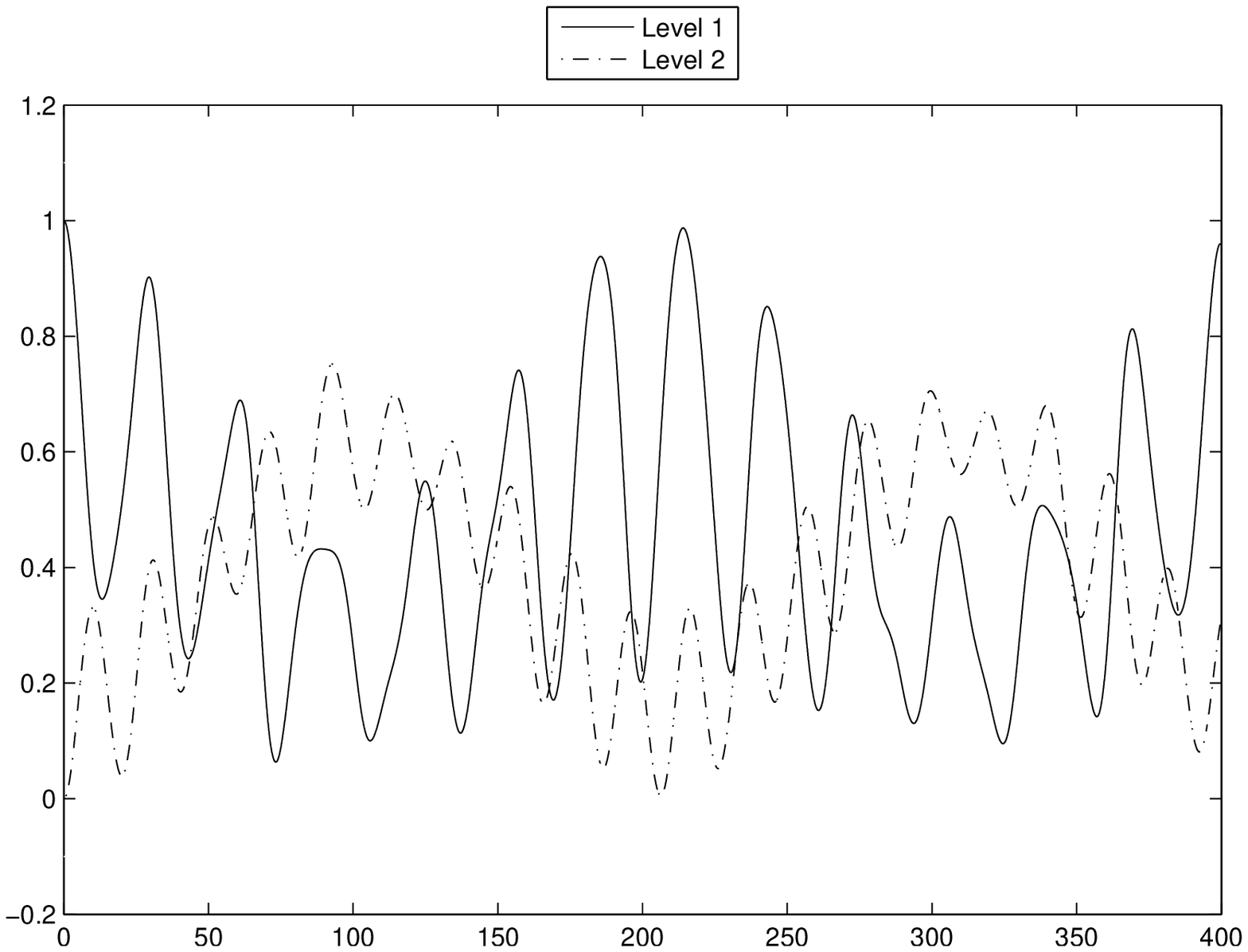}\hfill\\
\caption{\label{fig3b}\footnotesize Densities of the nutrients, $G_1$ and $G_2$, left, and of Levels 1 and 2, right. Initial conditions: nutrients and level 2 empty,
$G_1$, $G_2$ and level 1 completely filled.}
\end{center}
\end{figure}

Figure \ref{fig3b} describes the results obtained with the same choice of parameters and assuming that the nutrients and level 2 are empty at $t=0$, while
the two garbages and level 1 are completely filled.

\begin{figure}[h]
\begin{center}
\includegraphics[width=0.47\textwidth]{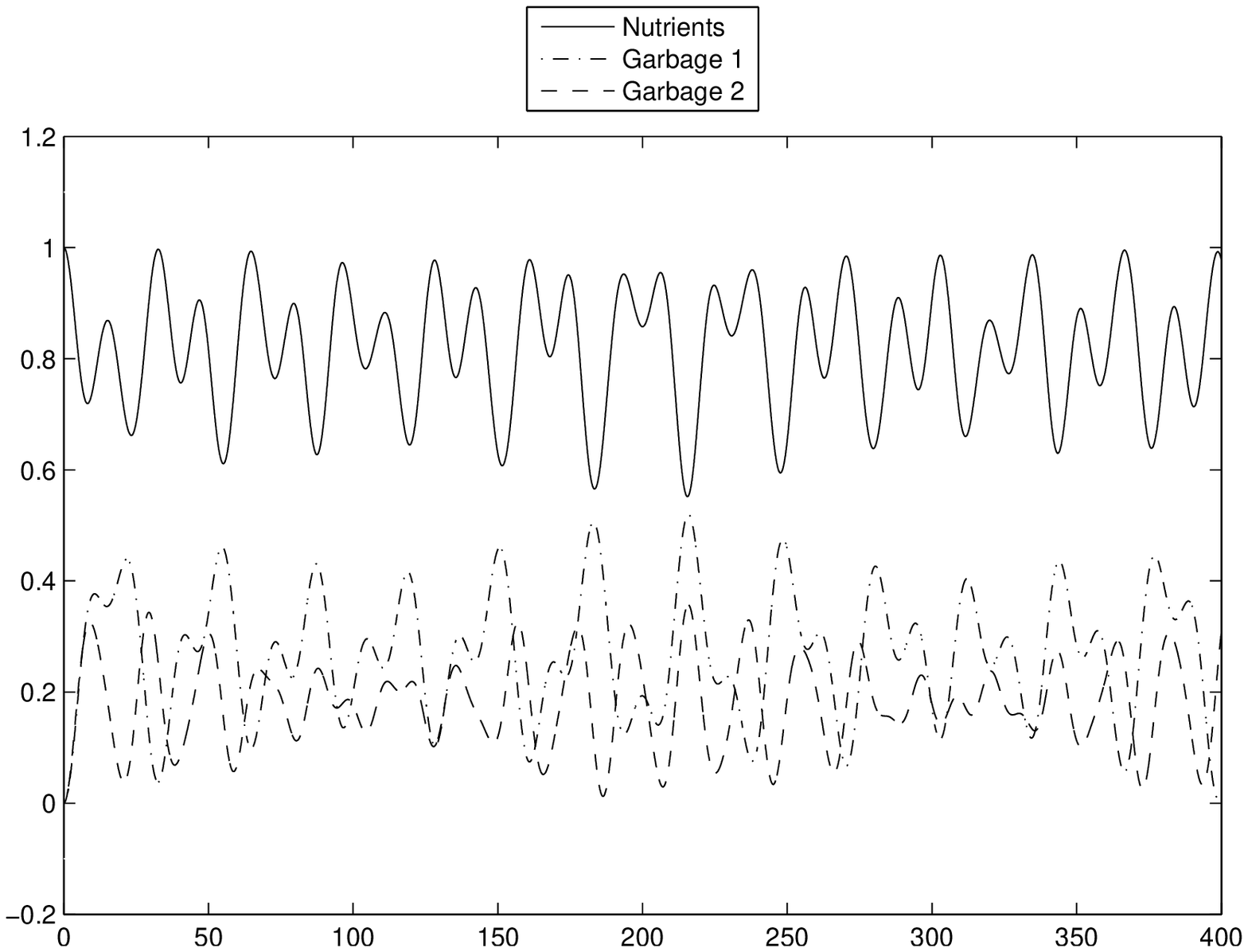}\hspace{8mm}
\includegraphics[width=0.47\textwidth] {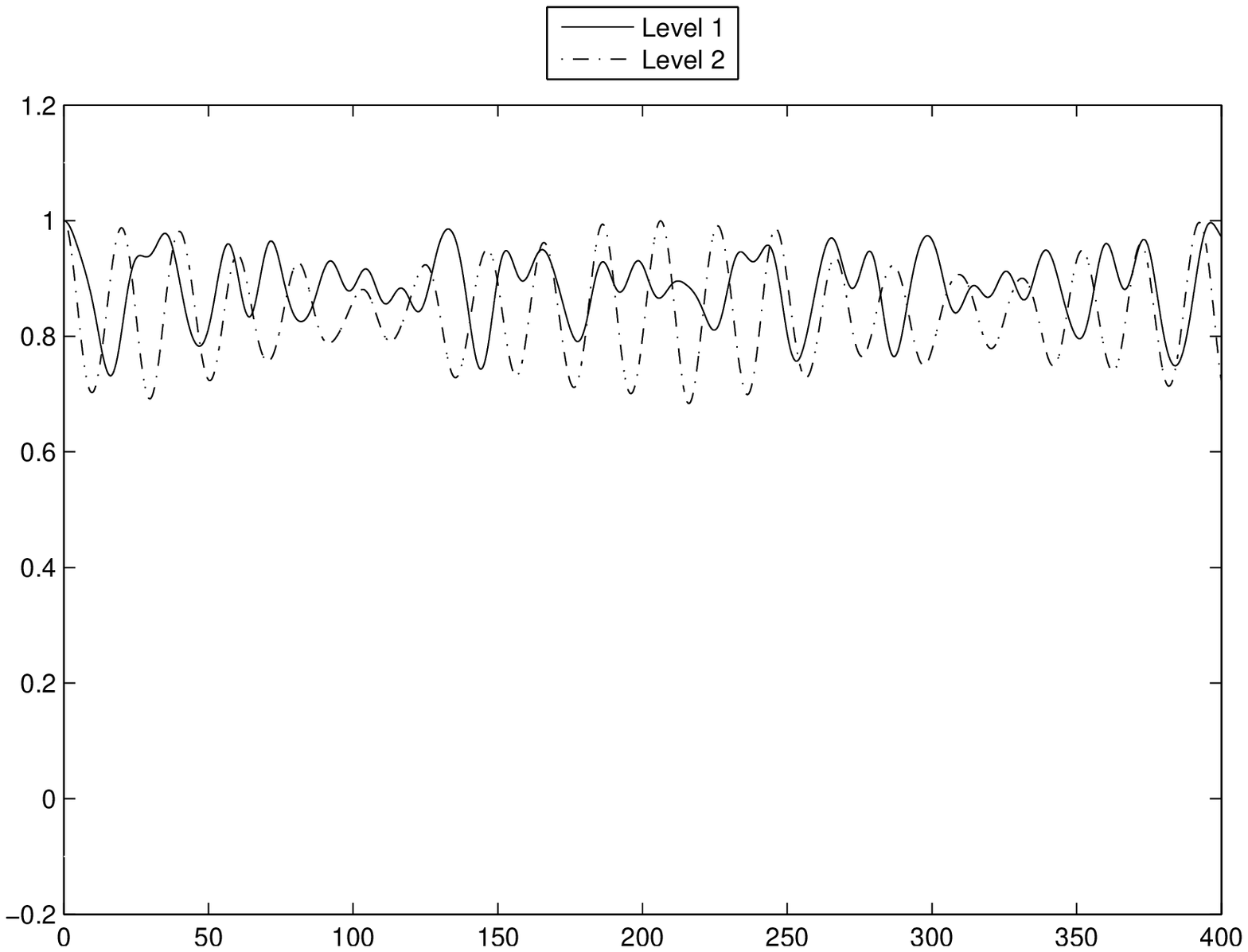}\hfill\\
\caption{\label{fig3c}\footnotesize Densities of the nutrients, $G_1$ and $G_2$, left, and of Levels 1 and 2, right. Initial conditions: nutrients, levels 1 and 2 completely filled,
$G_1$ and $G_2$ empty.}
\end{center}
\end{figure}

Figure \ref{fig3c} describes the results obtained with the same choice of parameters and assuming that the nutrients, levels 1 and 2 are completely filled at $t=0$, while
the two garbages are empty.

Among other things, these figures show that the fluctuations of $G_2$ are smaller than those of the other compartments. This is expected since the value
of the related $\omega$ in $H_0$ is the largest one, and it has been checked in several models (see Ref.~\onlinecite{bagbook}) that this parameter measures the
inertia of that particular ingredient of the system. As far as the densities of levels 1 and 2 are concerned, Figures \ref{fig3a} and \ref{fig3b} show that it happens that the
originally empty level acquires a density which is larger than the density of the other, originally filled, level. In other words, we see that
an inversion of the populations is possible. Figures \ref{fig3c} show that, starting with
a \emph{clean} initial condition (nutrients, levels 1 and 2 completely filled, and no waste products at all), the densities of the garbages remain reasonably low, while the nutrients and the
densities of the organisms in levels 1 and 2 oscillate around high values. This could be interesting for concrete applications, when one needs to maximize the {\em efficiency} of the ecosystem.
We also notice that no damping is allowed
within our present scheme. This is not surprising \cite{bagbook} due to the fact that no compartment here has an infinite number of degrees
of freedom. We will briefly come back on this aspect in Section \ref{sect5}.

\section{A nonlinear model with two garbages}
\label{sect4}

\begin{figure}[h]
\begin{center}
\includegraphics[width=0.47\textwidth]{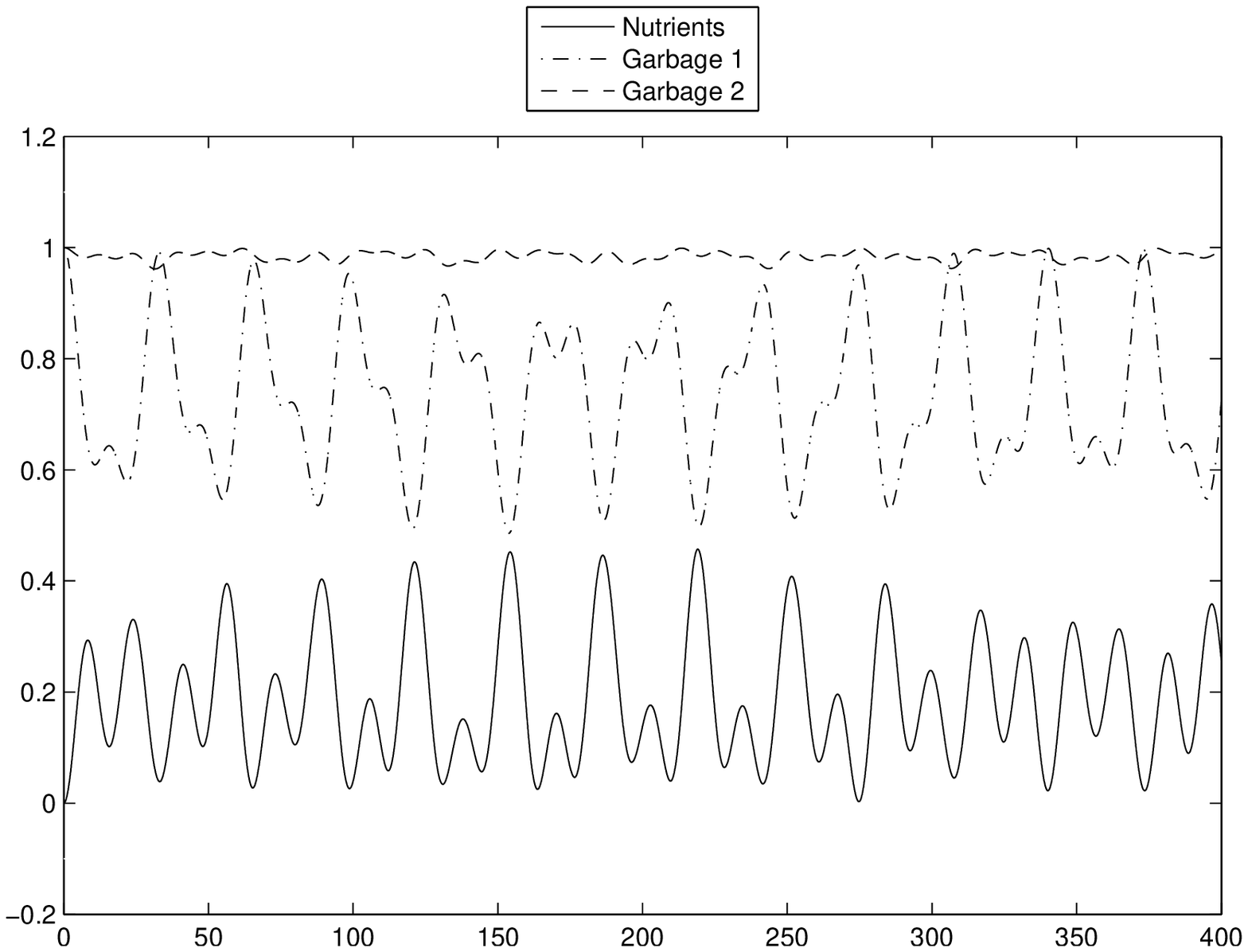}\hspace{8mm}
\includegraphics[width=0.47\textwidth]{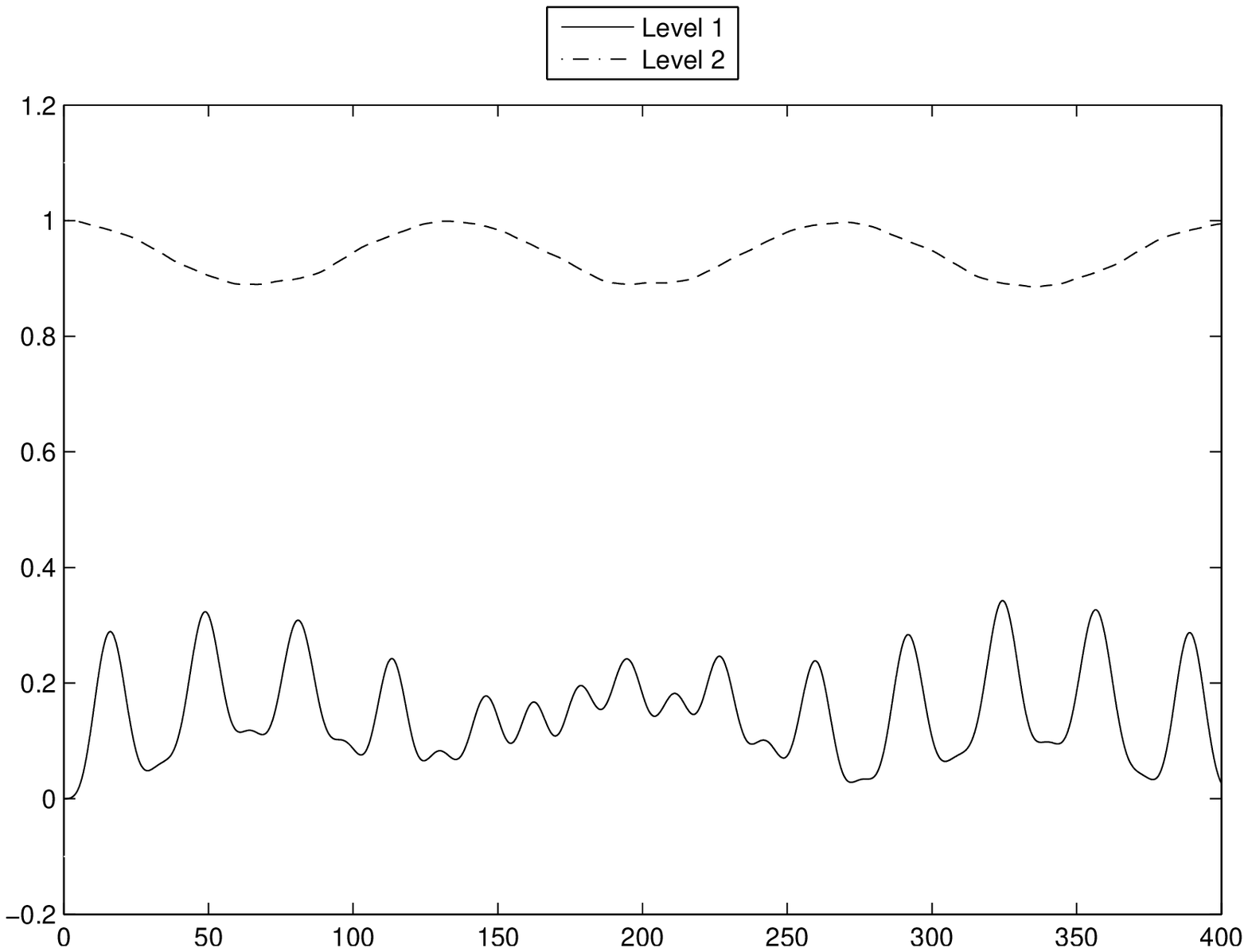}\hfill\\
\caption{\label{fig4a}\footnotesize Densities of the nutrients, $G_1$ and $G_2$, left, and of Levels 1 and 2, right. Initial conditions: nutrients and level 1 empty,
$G_1$, $G_2$ and level 2 completely filled.}
\end{center}
\end{figure}

The same system schematically described in Figure \ref{fig2} can be considered in a slightly different way: instead of considering two
different, quadratic terms, in $H$, to represent the interaction of the various levels with the two garbages, we could also consider a single
cubic contribution (see the first term in $H_I$ below). For instance, $a_j\,a_{N+1}^\dagger\,a_{N+2}^\dagger$ models the fact that the density
of the $j$-th level decreases while, simultaneously, the densities of both $G_1$ and $G_2$ increase: an organism through its metabolism or dying produce garbage of
two different kinds, soft and hard. The full hamiltonian of the system is the following one:
 \be
\left\{
\begin{aligned}
&H=H_0+ H_I, \qquad \hbox{ with }  \\
&H_0 = \sum_{j=0}^{N+2}\,\omega_{j}\, a_j^\dagger\,a_j,\\
&H_I= \sum_{j=1}^N\,\lambda_j\left(a_j\,a_{N+1}^\dagger\,a_{N+2}^\dagger+a_{N+2}\,a_{N+1}\,a_j^\dagger\right)\\
&\quad+\sum_{j=1}^2\,\nu^{(j)}\left(a_0\,a_{N+j}^\dagger+a_{N+j}a_0^\dagger\right)
 +\sum_{j=0}^{N-1}\,\nu_j\left(a_j\,a_{j+1}^\dagger+a_{j+1}a_j^\dagger\right).
\end{aligned}
\right.
\label{41}
\en
The notation is the same as before: for instance, zero is the fermionic mode for the nutrients, while $N+1$ and $N+2$ are the
modes for the two garbages. The physical interpretation of the hamiltonian is easily found:  $H_I$ describes an
interaction between the levels and the two garbages (first contribution), the nutrients and the two garbages (second contribution), and a {\em hopping}
term (third term): the nutrients are used to feed the organisms of level 1, and the organisms of level $j$ feed those of level $j+1$ ($j=1,\ldots,N-1$). The conjugate term,
$a_{j+1}a_j^\dagger$, is needed in order to render the hamiltonian self-adjoint,
since all the parameters are supposed here to be real. The Heisenberg equations of motion look much harder than the previous ones. Indeed, calling $X:=\sum_{l=1}^N\lambda_l\,a_l$, we
have
\be
\left\{
\begin{aligned}
&\dot a_0=i\left(-\omega_0a_0+\nu_0a_1+2Xa_0a_{N+1}^\dagger a_{N+2}^\dagger+2a_{N+2}a_{N+1}X^\dagger a_0+\nu^{(1)}a_{N+1}+\nu^{(2)}a_{N+2}\right),  \\
&\dot a_j=i\left(-\omega_ja_j+\nu_ja_{j+1}+\nu_{j-1}a_{j-1}+2Xa_ja_{N+1}^\dagger a_{N+2}^\dagger+ a_{N+2}a_{N+1}(2X^\dagger a_j-\lambda_j\Id)\right),\\
&\dot a_N=i\left(-\omega_Na_N+\nu_{N-1}a_{N-1}+2Xa_Na_{N+1}^\dagger a_{N+2}^\dagger+a_{N+2}a_{N+1}(2X^\dagger a_N-\lambda_N\Id)\right),\\
&\dot a_{N+1}=i\left(-\omega_{N+1}a_{N+1}+Xa_{N+2}^\dagger(\Id-2a_{N+1}^\dagger a_{N+1})+\nu^{(1)}a_0\right),\\
&\dot a_{N+2}=i\left(-\omega_{N+2}a_{N+2}+Xa_{N+1}^\dagger(2a_{N+2}^\dagger a_{N+2}-\Id)+\nu^{(2)}a_0\right),
\end{aligned}
\right.
\label{42}
\en
$l=1,2,\ldots,N-1$. It is evident that this system is not closed. In order to close it, we have to consider also the
hermitian conjugate of these equations. In this way we get a nonlinear system, whose solution can be found numerically. Notice also that,
because of the nonlinearity, the operator $N_{tot}$ introduced in Section \ref{sect3} does not commute with the hamiltonian, $[H, N_{tot}]\neq 0$, and it is not
evident if any other integral of motion exists at all. Losing the linearity looks like {\em opening the system} to the outer world: part of $N_{tot}$
could be lost or created, during the time evolution. This could have interesting consequences, since we might expect that a realistic ecosystem is
not entirely closed. On a mechanical level, this looks like having a sort of unavoidable friction in the system, friction which can be made
small, or even very small, but not zero. However, the plots in Figure \ref{fig4a}, which are produced fixing, as in Section 3, $N=2$, do not show any
clear damping effect, and in fact this will be introduced phenomenologically in Section
\ref{sect5}, by adding a small imaginary part to some
parameter involved in the hamiltonian.

These plots are produced with the following choice of the parameters involved in the hamiltonian, choice which relies on the same arguments given in Section \ref{sect3}: $\omega_0=0.05$, $\omega_1=0.1$, $\omega_2=0.2$,
$\omega_3=0.3$, $\omega_4=0.45$, $\lambda_1=0.005$,  $\lambda_2=0.009$, $\nu_0=0.1$, $\nu_1=0.01$, $\nu^{(1)}=0.1$ and $\nu^{(2)}=0.03$.  Notice that $\nu^{(1)}$ is taken larger than $\nu^{(2)}$ since $G_1$ is assumed to produce nutrients more quickly than $G_2$.

\begin{figure}[h]
\begin{center}
\includegraphics[width=0.47\textwidth]{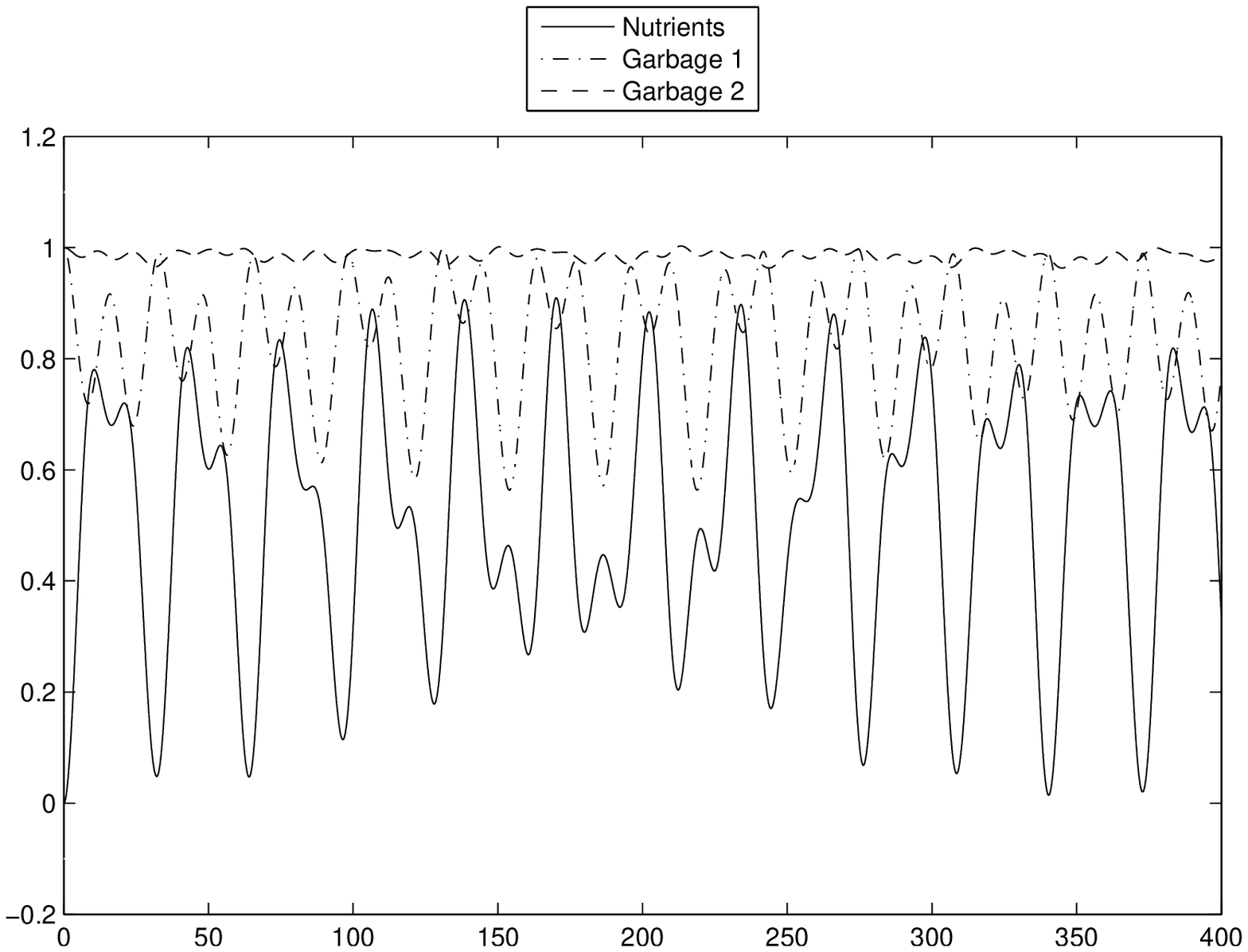}\hspace{8mm}
\includegraphics[width=0.47\textwidth]{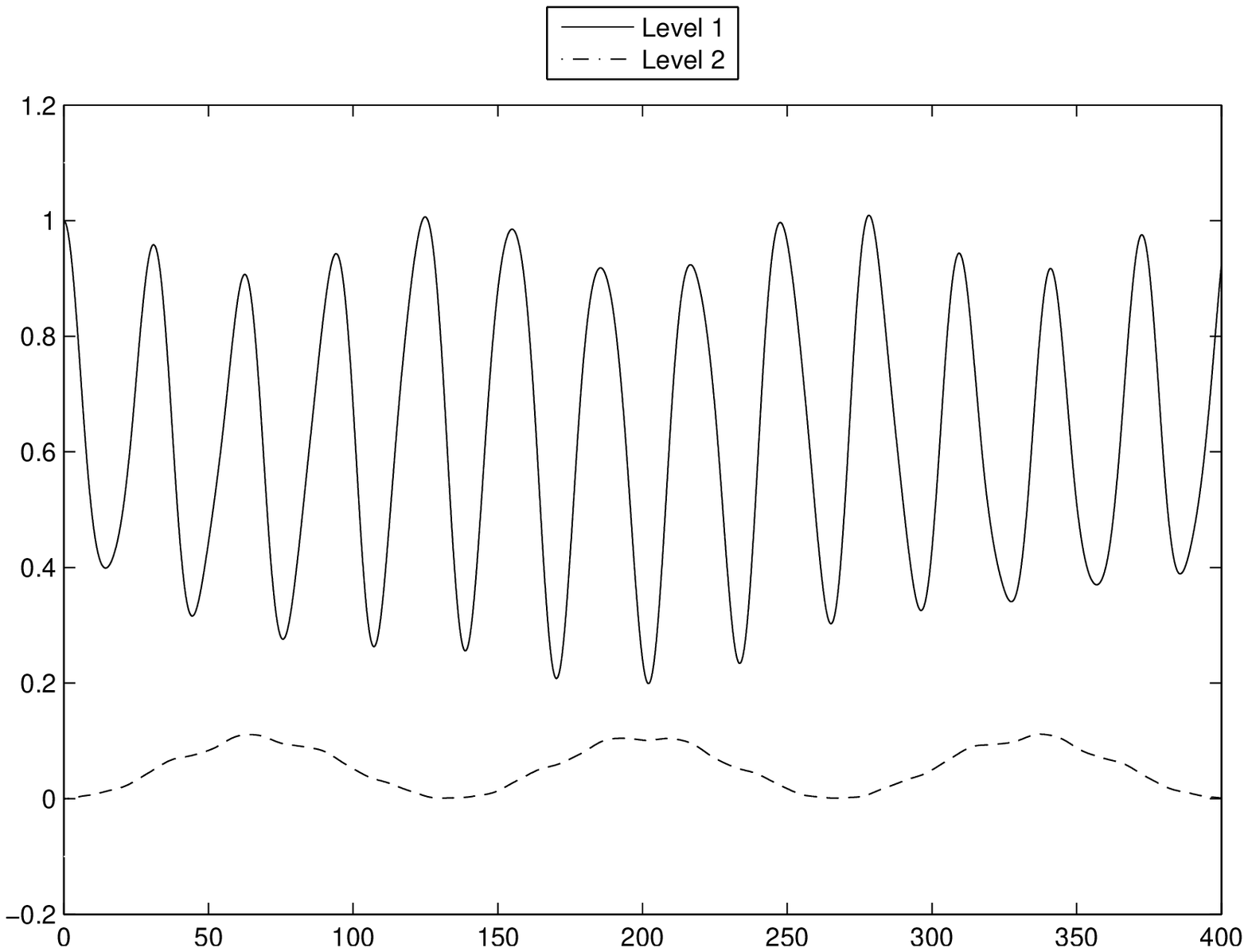}\hfill\\
\caption{\label{fig4b}\footnotesize Densities of the nutrients, $G_1$ and $G_2$, left, and of Levels 1 and 2, right. Initial conditions: nutrients and level 2 empty,
$G_1$, $G_2$ and level 1 completely filled.}
\end{center}
\end{figure}

Figure \ref{fig4a} shows, among other things, that level 2 and $G_2$ change in time less than the other compartments, as expected. The nutrients and
$G_1$ appear to be exactly out of phase. This is interesting, since it suggests that the soft garbage turns into nutrients quite easily (actually,
simultaneously), while the hard garbage in almost not involved into this transformation. During the time evolution, the density of level 1 can
change of, at most, the $30\%$ of its initial value, while the density of level 2 can decrease, at most, of the $10\%$. Similar features are
depicted in Figure \ref{fig4b}, which differ from the previous one only for the initial conditions.

Again we see the effect of the inertia which makes the second level and $G_2$ almost constant in time, especially when compared with the other
compartments. This time the nutrients and $G_1$ are no longer exactly out of phase, as it is probably more realistic: the garbage in $G_1$ does not
turn into nutrients instantaneously. It takes some time.

\begin{figure}[h]
\begin{center}
\includegraphics[width=0.47\textwidth]{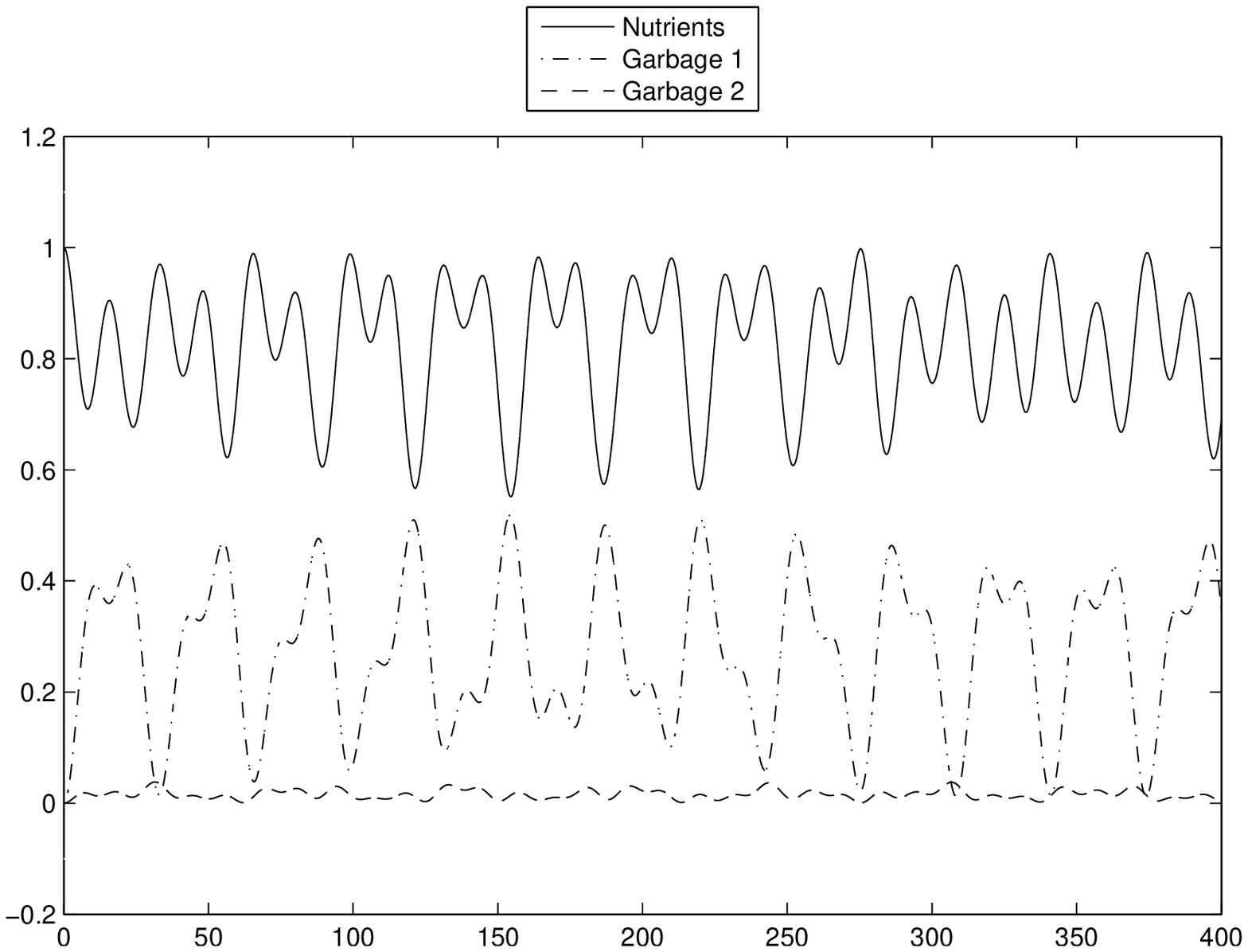}\hspace{8mm}
\includegraphics[width=0.47\textwidth]{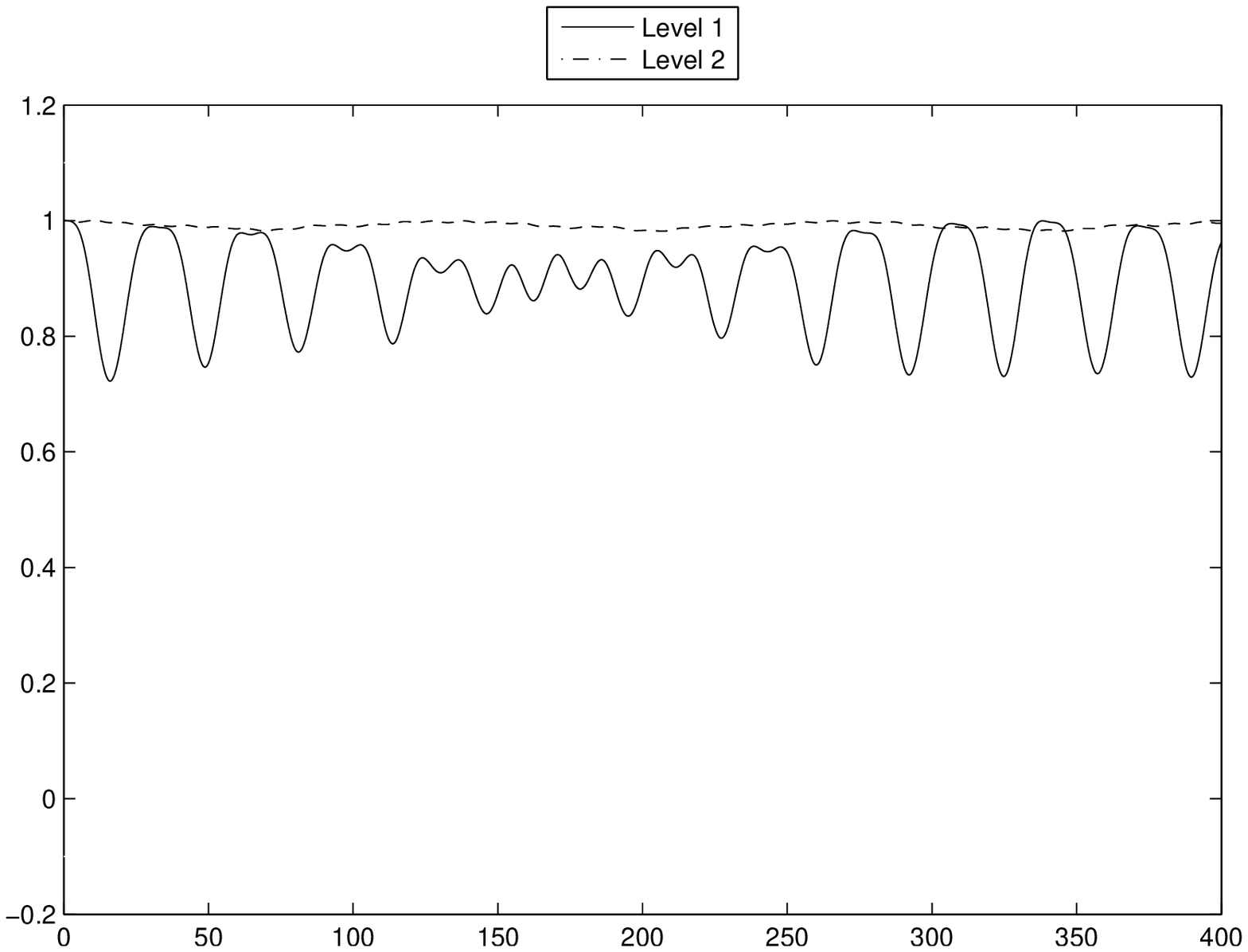}\hfill\\
\caption{\label{fig4c}\footnotesize Densities of the nutrients, $G_1$ and $G_2$, left, and of Levels 1 and 2, right. Initial conditions: nutrients, levels 1 and 2 completely filled,
$G_1$ and $G_2$ empty.}
\end{center}
\end{figure}

In Figure \ref{fig4c}, starting with an initial consition with no waste products, we observe that organisms of level 2 undergo negligible variations, nutrients and organisms of level 1 oscillate around very high values, whereas waste compartments do not assume
high values of the densities (garbage 2 remains almost empty).

However, we see again that no evident damping takes place in this model, at least with our choice
of the parameters (as well as with other choices we have considered that are not reported here). The presence of damping will be discussed in Section \ref{sect5}.

\section{A phenomenological damping}
\label{sect5}

As widely discussed in Ref.~\onlinecite{bagbook} and references therein, a rigorous way to describe damping in a quantum system is to {\em open it},
making the system to interact with a suitable reservoir. In this way the dynamics
of the full system remains unitary, even if an exchange between the system and
the (infinitely extended) reservoir allows us to describe quantities which are not conserved during the time evolution. However, quite often,
an {\em effective approach} is used, \emph{i.e.}, that of replacing self-adjoint with non self-adjoint hamiltonians, keeping unchanged the other rules
of the game. In particular, as shown in  Appendix B, in order to describe a damping effect, it is sufficient (but not rigorous, we should say)
to replace some real parameters involved in the hamiltonian with complex numbers. This is exactly what we will do in this section. In particular, we will
show that it is enough to add a small negative imaginary part to just a single parameter of $H_0$ in (\ref{31}) and (\ref{41}), to induce a
damping for all the compartments of our ecosystem.

\subsection{The linear case}

The numerical values of the parameters are exactly those of Section \ref{sect3}, except for $\omega_3$, which is no longer $\omega_3=0.3$ but it
is replaced now by $\omega_3=0.3-0.01\,i$: we are adding a negative and relatively small imaginary part to $\omega_3$. The reason for the choice of the sign of the imaginary part is suggested in Appendix B, where we show that we are forced to make a similar choice to get damping: taking a positive imaginary part for
$\omega_3$ produces a blow up of the solution!
As a matter of fact, this could also be deduced directly from the equations of motion.
To illustrate this, let us consider the differential equation $\dot x=-i\omega x$, where
$\omega=\omega_r+i\omega_i$, $\omega_r,\omega_i\in\mathbb{R}$, $x\in\mathbb{C}$.
It is clear that $x(t)=\exp(-i\omega t)x(0)=\exp(-i\omega_r t)\exp(\omega_i t)x(0)$,
which is decaying only if $\omega_i<0$.

\begin{figure}[h]
\begin{center}
\includegraphics[width=0.47\textwidth]{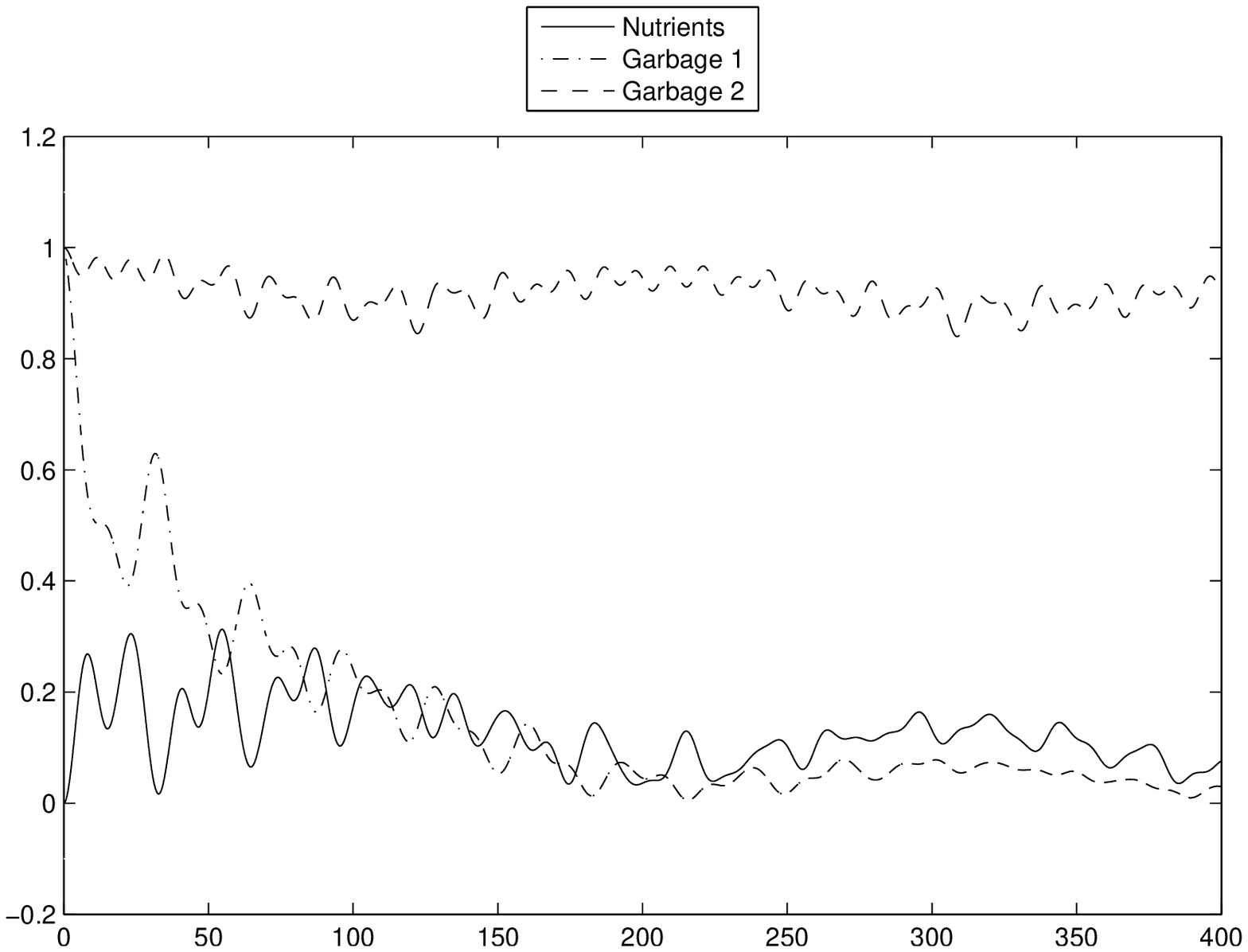}\hspace{8mm}
\includegraphics[width=0.47\textwidth]{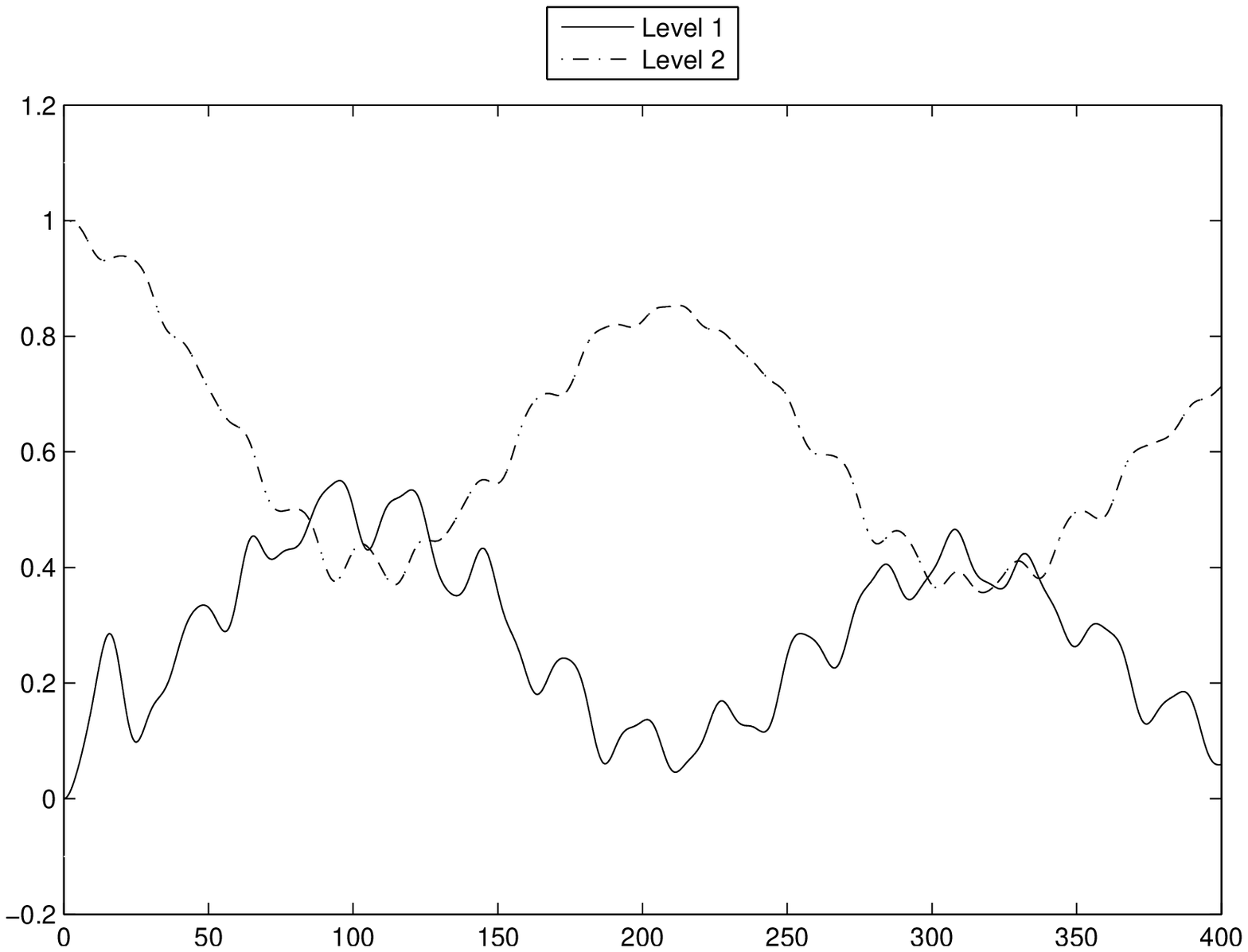}\hfill\\
\caption{\label{fig5a}\footnotesize Densities of the nutrients, $G_1$ and $G_2$, left, and of Levels 1 and 2, right. Initial conditions: nutrients and level 1 empty,
$G_1$, $G_2$ and level 2 completely filled.}
\end{center}
\end{figure}

\begin{figure}[h]
\begin{center}
\includegraphics[width=0.47\textwidth]{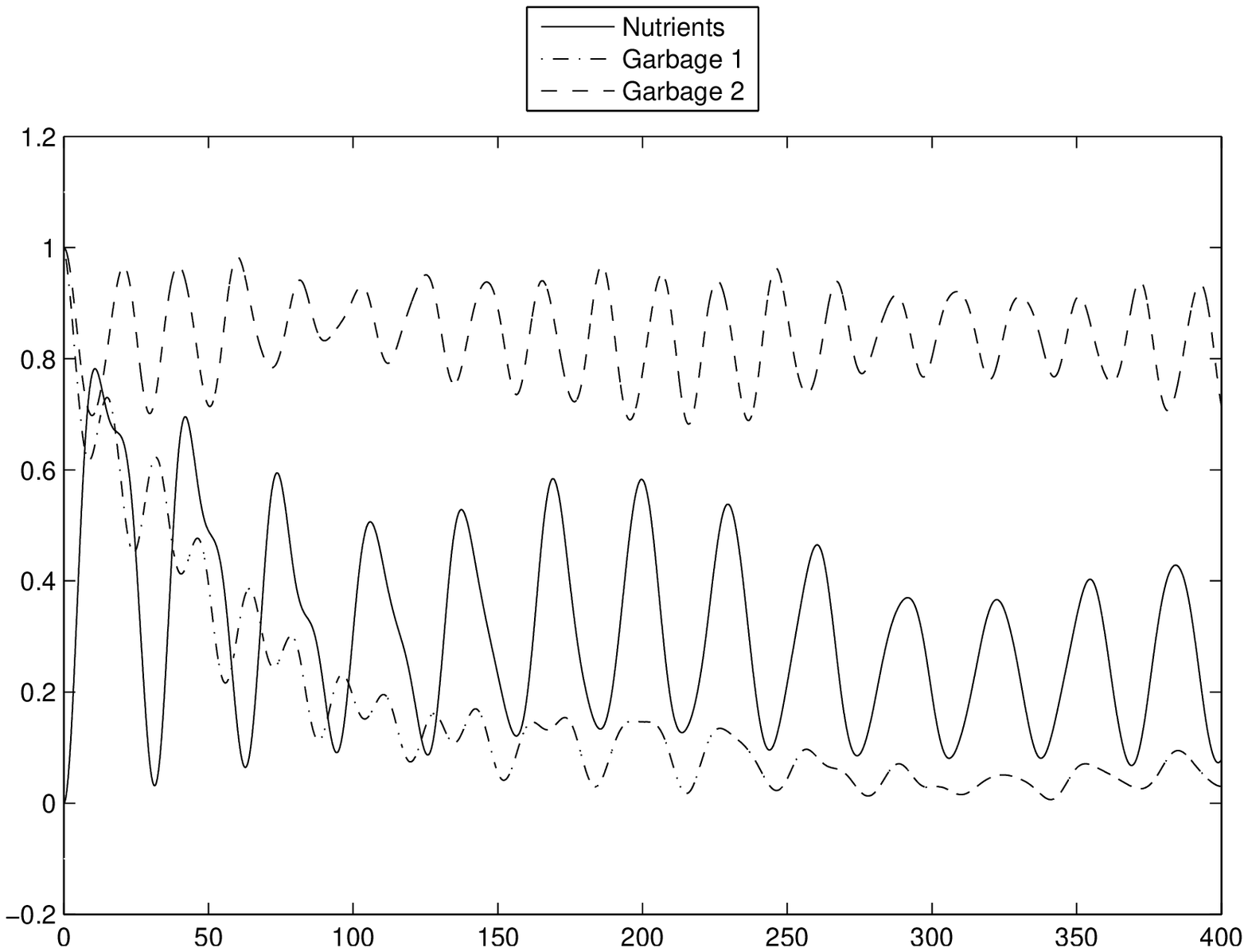}\hspace{8mm}
\includegraphics[width=0.47\textwidth]{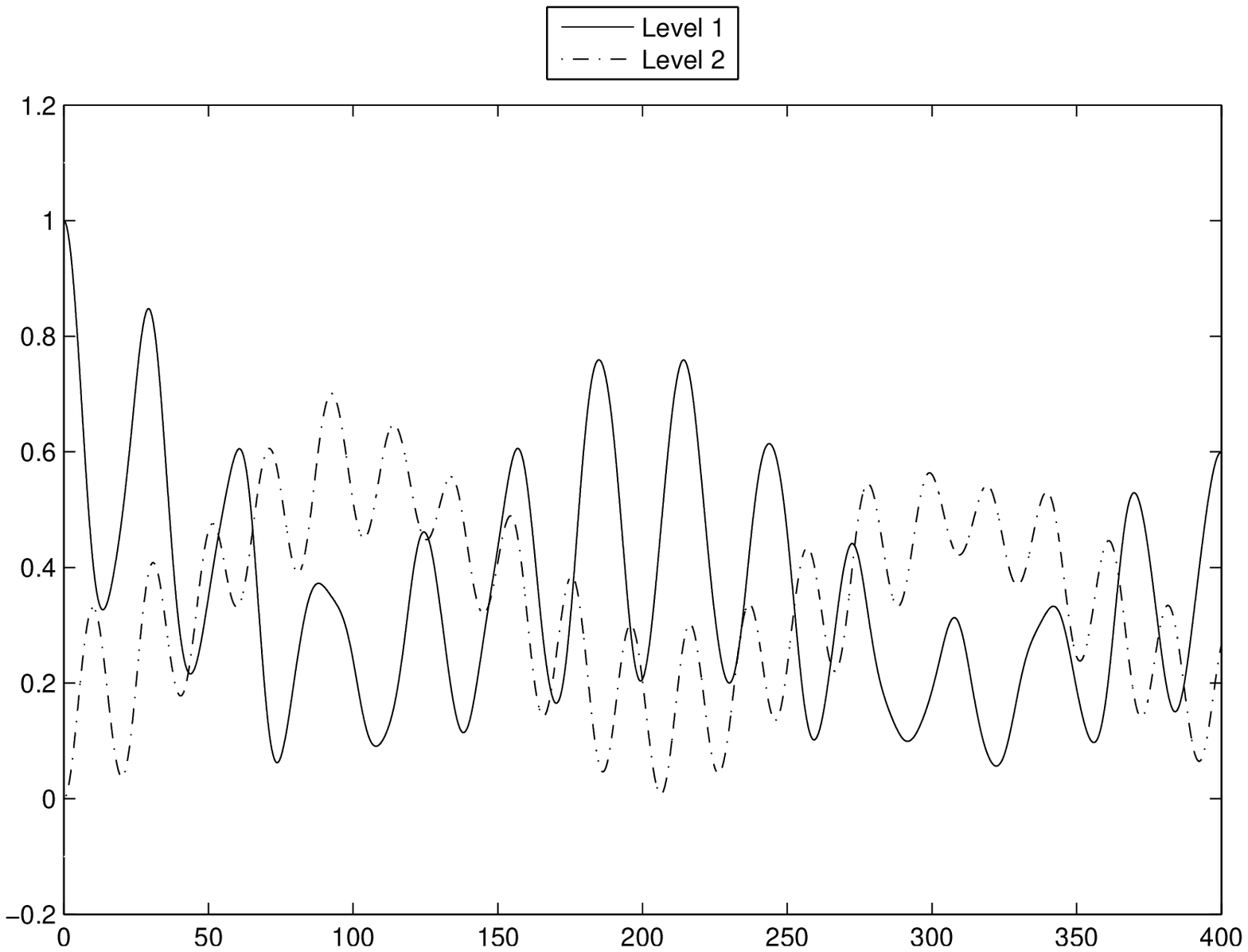}\hfill\\
\caption{\label{fig5b}\footnotesize Densities of the nutrients, $G_1$ and $G_2$, left, and of Levels 1 and 2, right. Initial conditions: nutrients and level 2 empty,
$G_1$, $G_2$ and level 1 completely filled.}
\end{center}
\end{figure}

\begin{figure}[h]
\begin{center}
\includegraphics[width=0.47\textwidth]{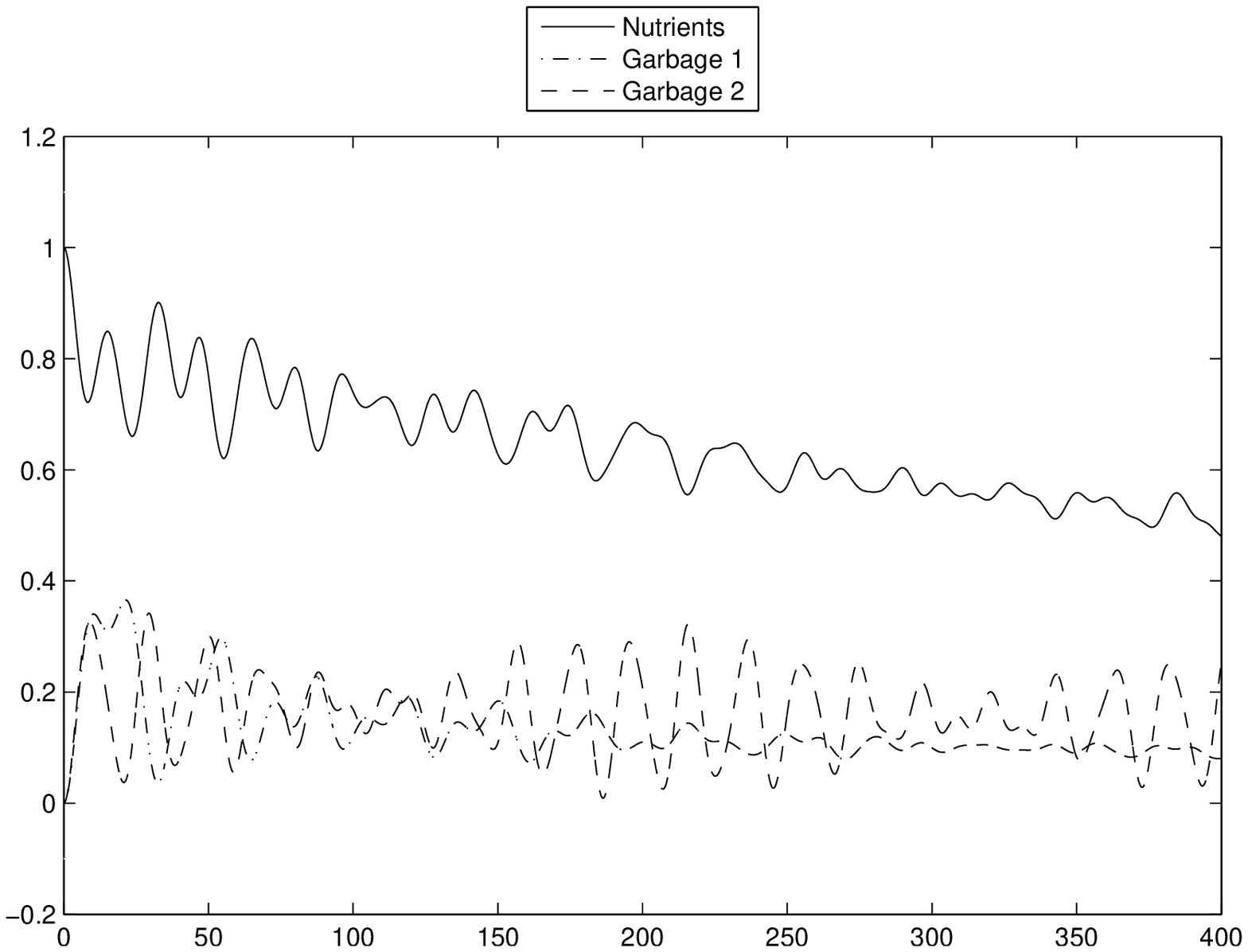}\hspace{8mm}
\includegraphics[width=0.47\textwidth] {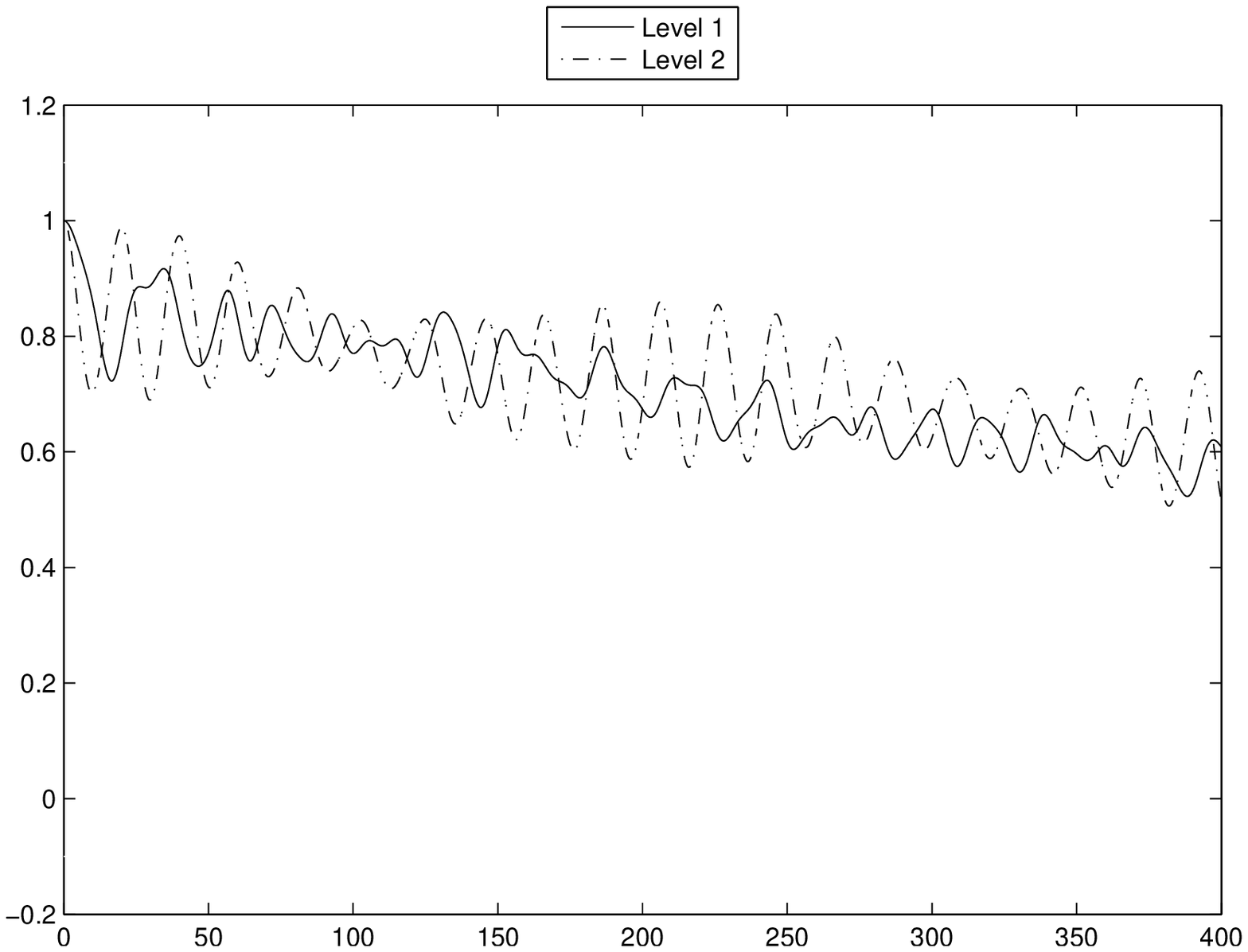}\hfill\\
\caption{\label{fig5c}\footnotesize Densities of the nutrients, $G_1$ and $G_2$, left, and of Levels 1 and 2, right. Initial conditions: nutrients, levels 1 and 2 completely filled,
$G_1$ and $G_2$ empty.}
\end{center}
\end{figure}

Choosing the initial conditions as in Figures \ref{fig3a}, \ref{fig3b}  and \ref{fig3c}, we get the plots shown in Figure \ref{fig5a}, \ref{fig5b} and \ref{fig5c}.

The damping is evident, especially in $G_1$. This is not surprising, since we have added the negative imaginary part exactly to the parameter
measuring the inertia of $G_1$. On the other hand, the decay of the density of $G_2$ looks rather slow. Again, this is what we expect because
of the large value of $\omega_4$, which makes the inertia of $G_2$ rather large.

\subsection{The nonlinear case}

\begin{figure}[h]
\begin{center}
\includegraphics[width=0.47\textwidth]{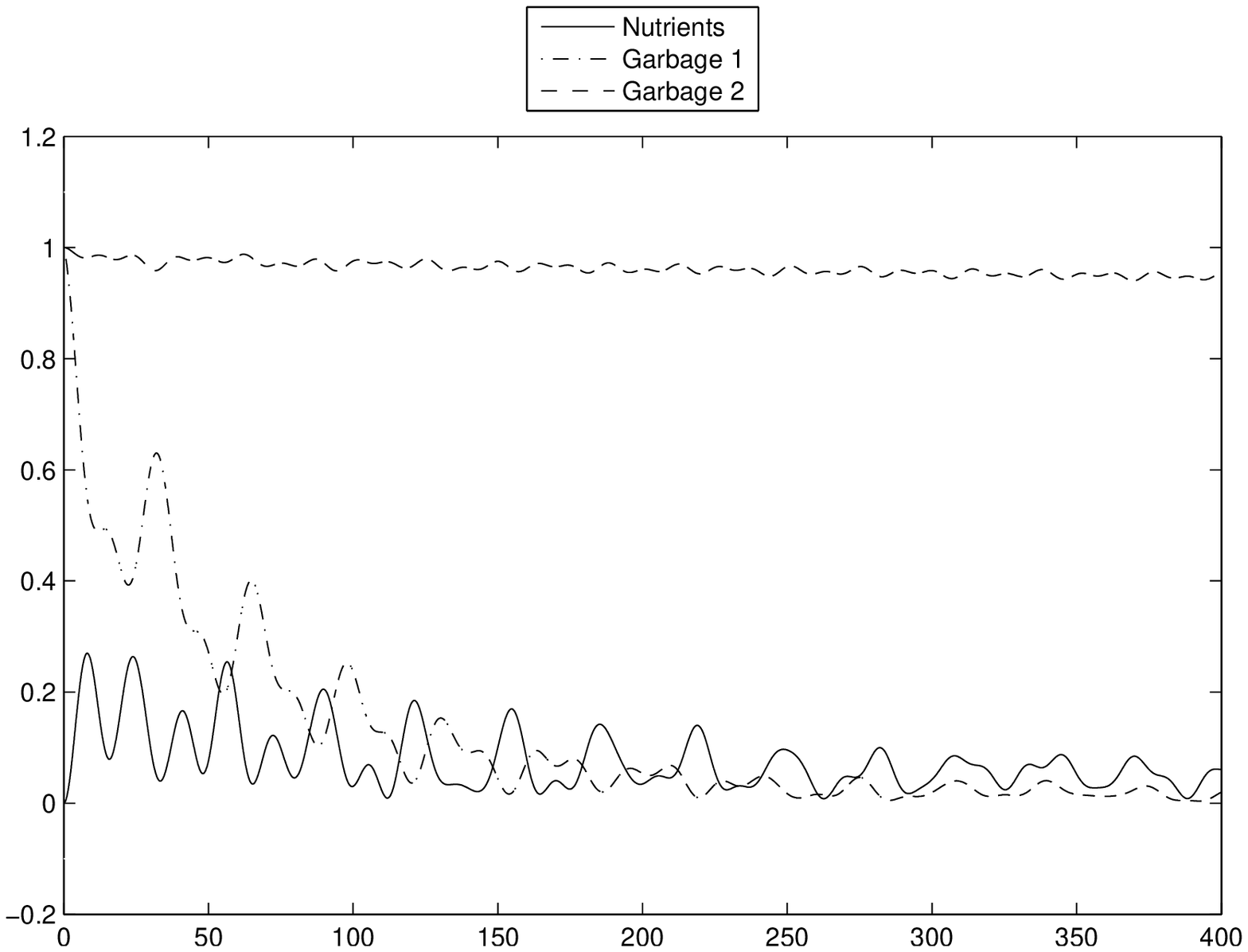}\hspace{8mm}
\includegraphics[width=0.47\textwidth]{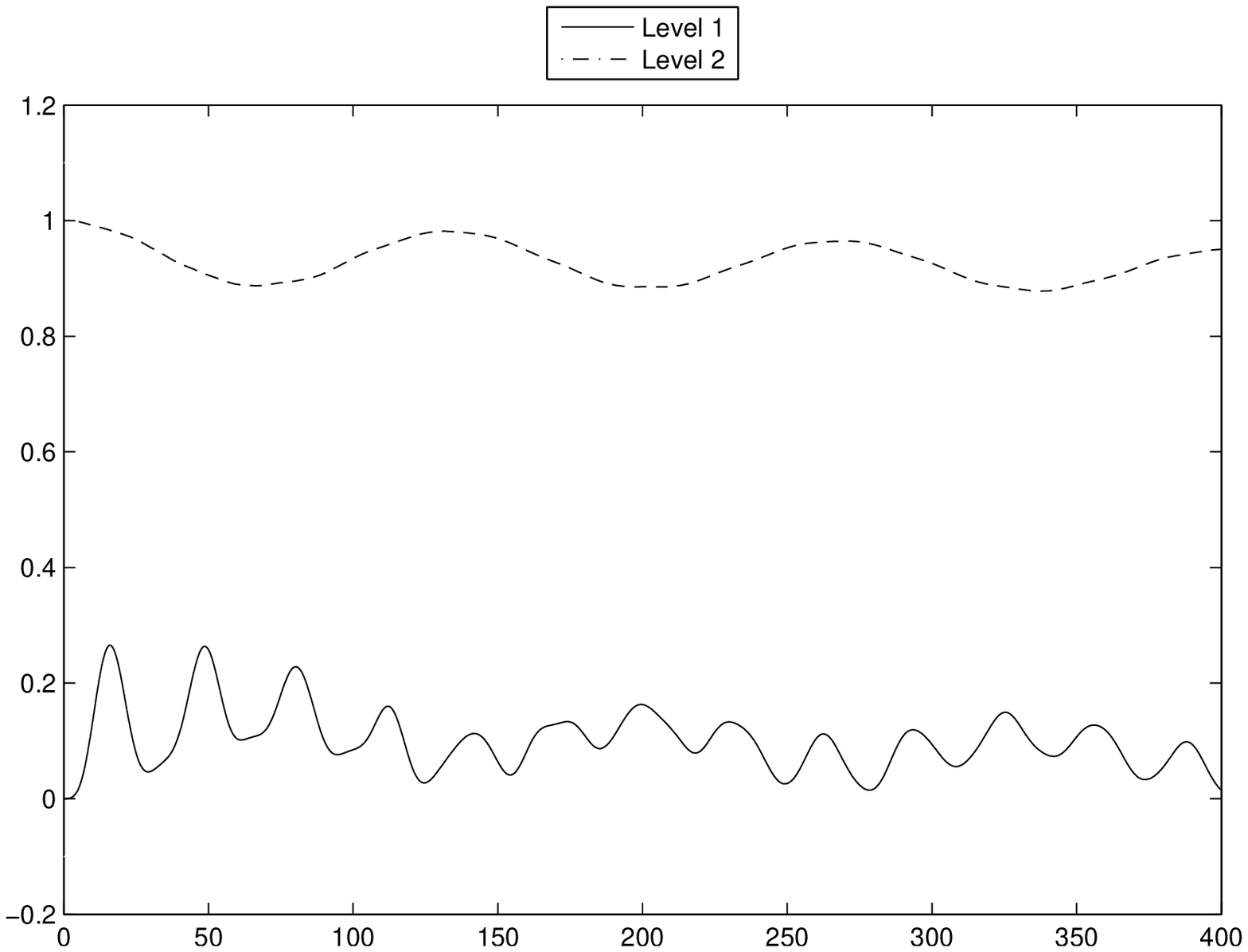}\hfill\\
\caption{\label{fig6a}\footnotesize Densities of the nutrients, $G_1$ and $G_2$, left, and of Levels 1 and 2, right. Initial conditions: nutrients and level 1 empty,
$G_1$, $G_2$ and level 2 completely filled.}
\end{center}
\end{figure}

\begin{figure}[h]
\begin{center}
\includegraphics[width=0.47\textwidth]{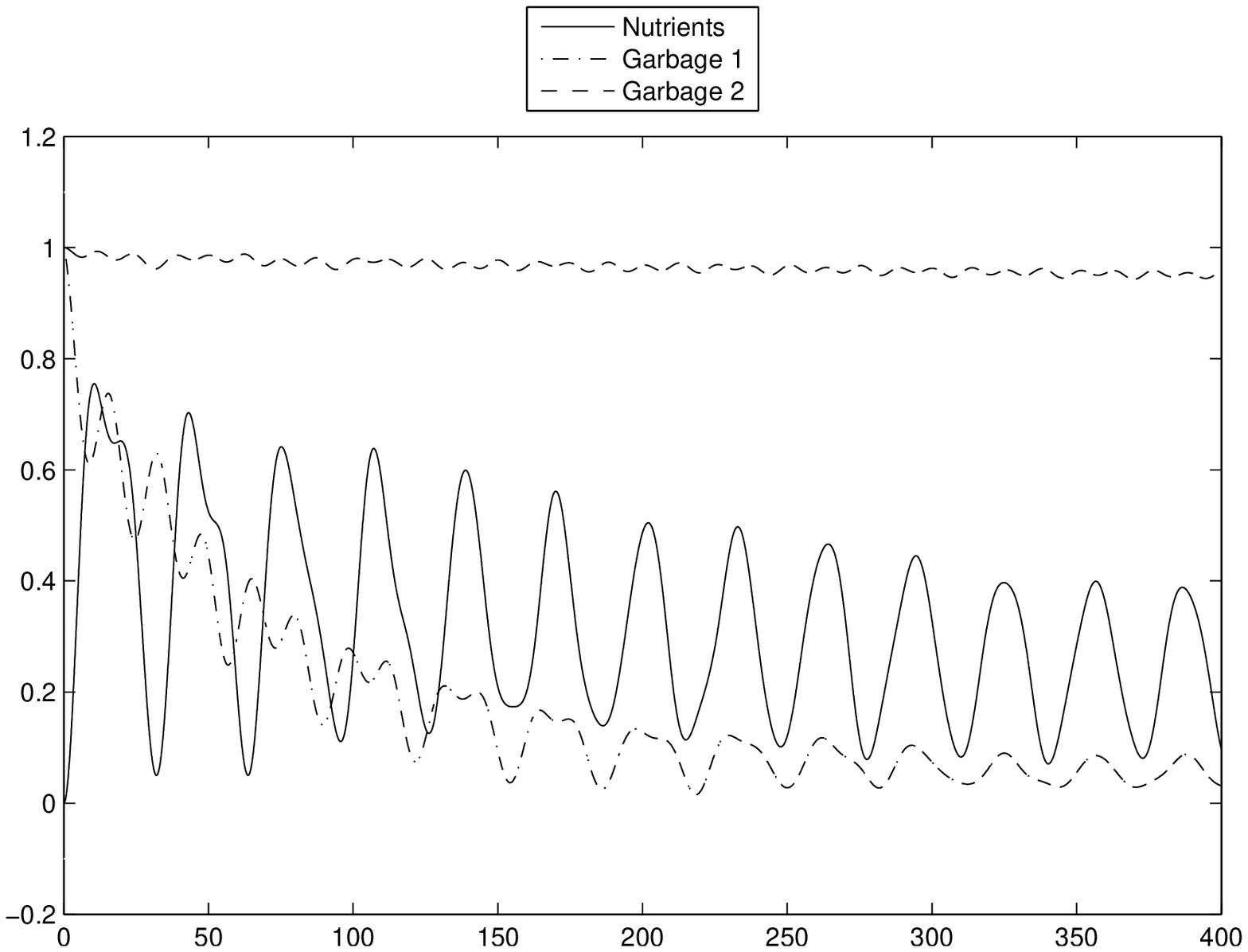}\hspace{8mm}
\includegraphics[width=0.47\textwidth]{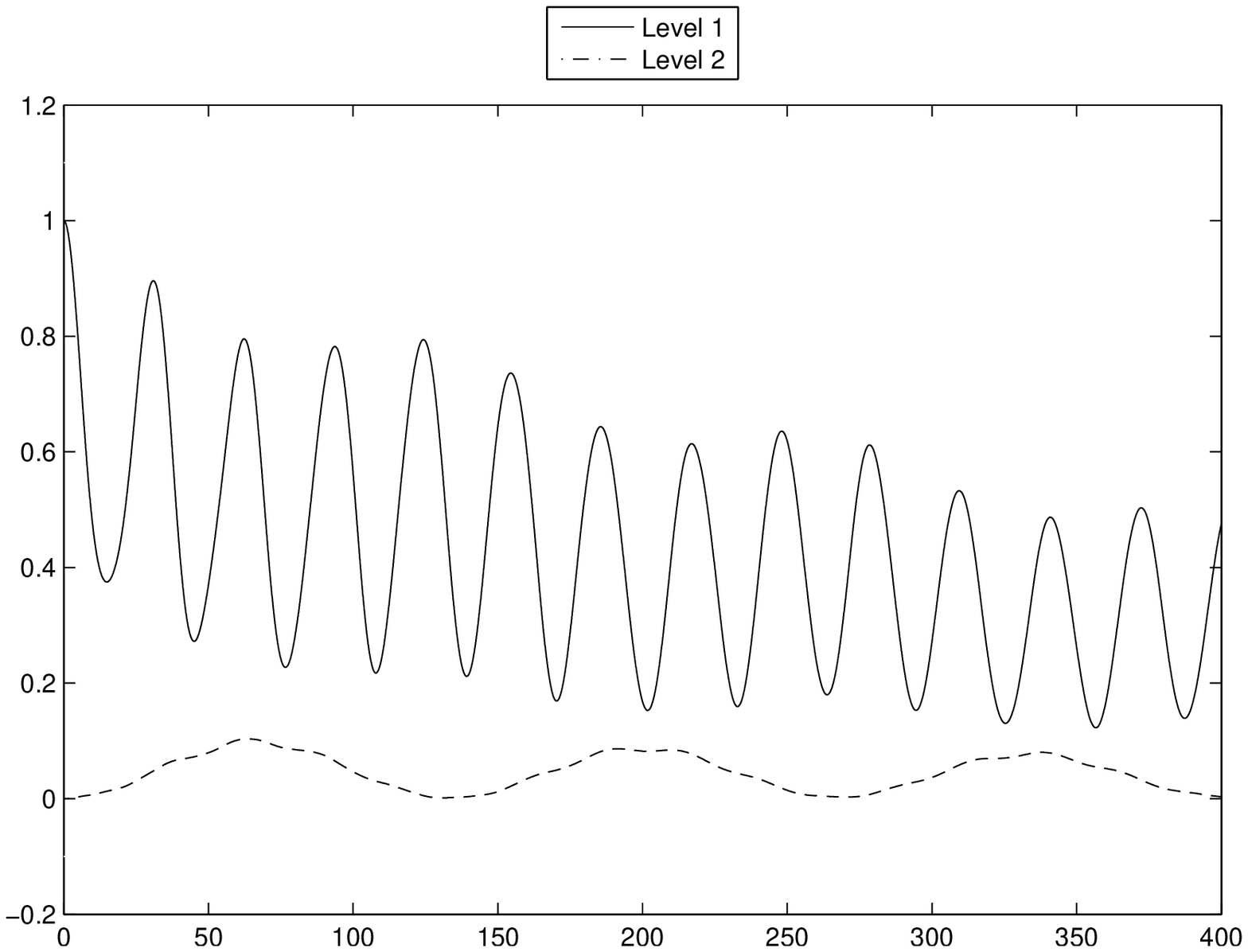}\hfill\\
\caption{\label{fig6b}\footnotesize Densities of the nutrients, $G_1$ and $G_2$, left, and of Levels 1 and 2, right. Initial conditions: nutrients and level 2 empty,
$G_1$, $G_2$ and level 1 completely filled.}
\end{center}
\end{figure}

\begin{figure}[h]
\begin{center}
\includegraphics[width=0.47\textwidth]{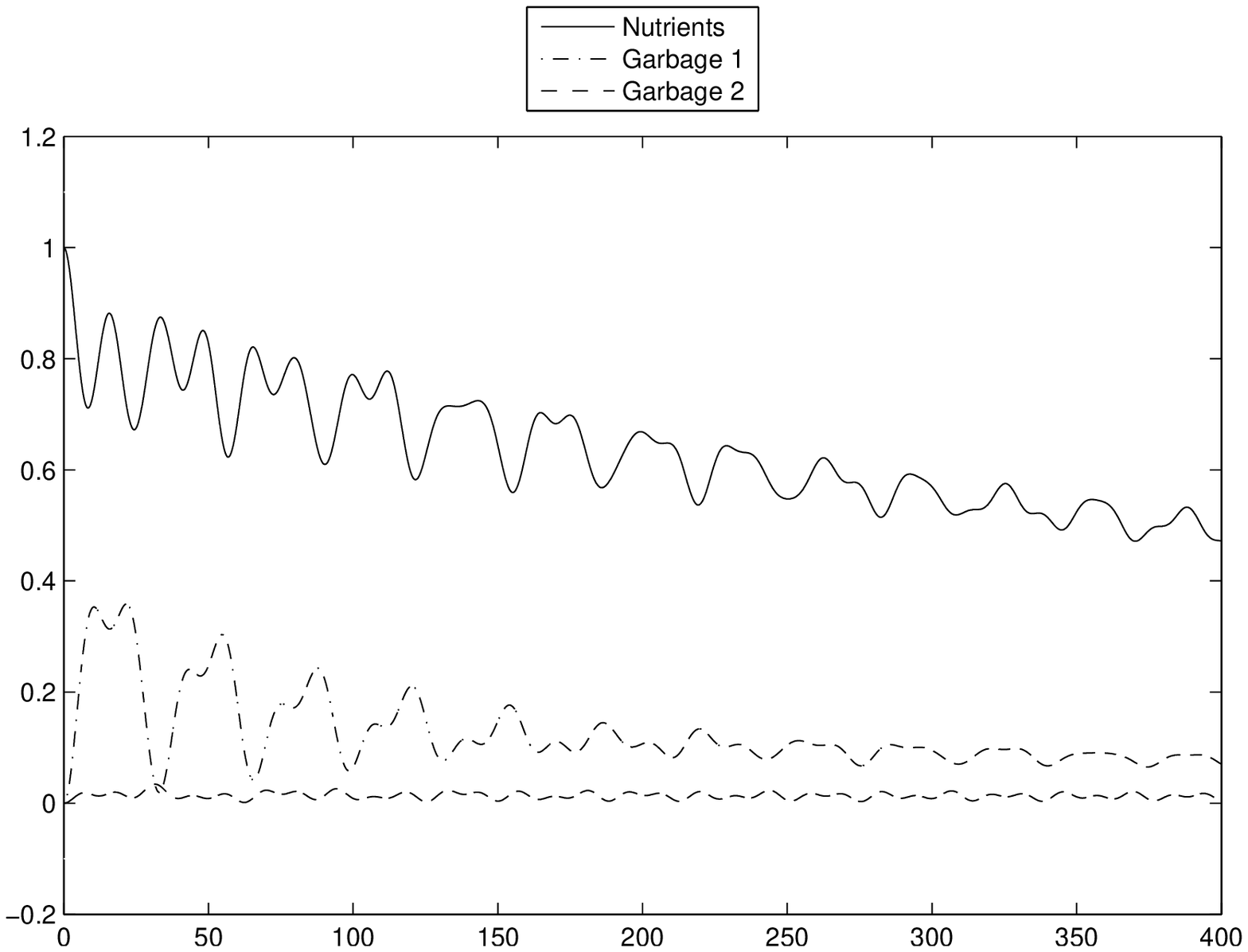}\hspace{8mm}
\includegraphics[width=0.47\textwidth]{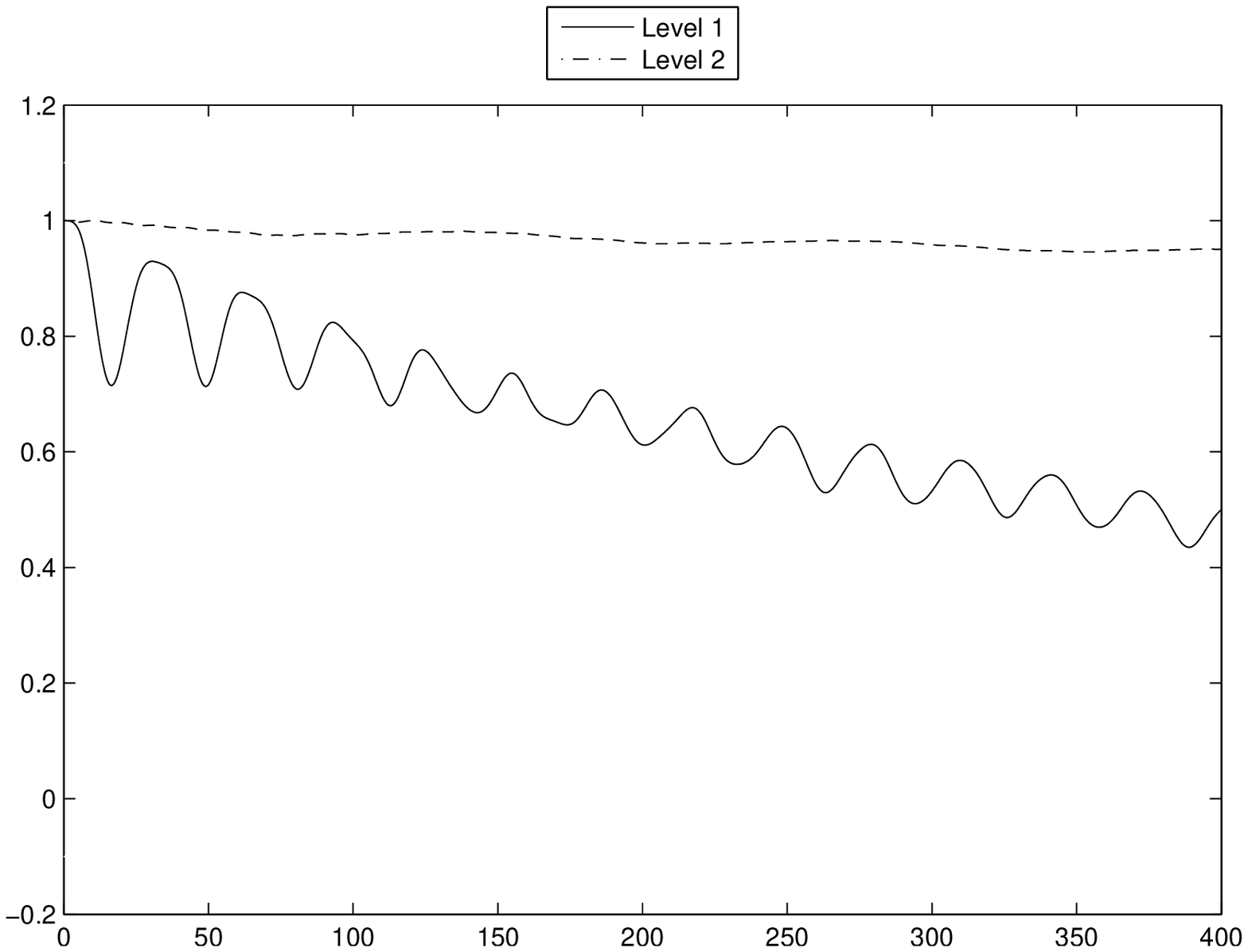}\hfill\\
\caption{\label{fig6c}\footnotesize Densities of the nutrients, $G_1$ and $G_2$, left, and of Levels 1 and 2, right. Initial conditions: nutrients and level 2 empty,
$G_1$, $G_2$ and level 1 completely filled.}
\end{center}
\end{figure}

The numerical values of the parameters are exactly those of Section \ref{sect4}, except for $\omega_3$, which again is no longer $\omega_3=0.3$ but,
as for the linear case, is now replaced by $\omega_3=0.3-0.01\,i$. In analogy with what we have already discussed, we observe that numerical computations show that if we take $\Im(\omega_3)>0$, even very small, rather than damping we get the blow up of the
densities. Figures \ref{fig6a}, \ref{fig6b} and \ref{fig6c} should be compared with Figures \ref{fig5a}, \ref{fig5b} and \ref{fig5c}, which
are deduced exactly with the same initial conditions and, except for $\Im(\omega_3)$, with the same values of the parameters of $H$.

In all these figures the damping effect is again clearly shown. However, due to the inertia of, say, $G_2$, we see that its density decreases much
slower than, say, that of $G_1$. The decay of level 2 is also much slower than that of level 1, because of their different inertia parameters.

These cases show that damping can be deduced adopting a standard Heisenberg-like dynamics, simply by replacing the original real
parameters in $H$ with some complex quantities. Moreover, what we have shown here, is that it is not important, or necessary, to replace all
the real parameters of $H$ with complex quantities. For instance, all the interaction parameters need not to be changed. By allowing  only one parameter of the free hamiltonian $H_0$ to be complex (with a negative imaginary part), we are able to get damping for the densities of
all the compartments: this is  exactly the behavior we expect in a realistic closed ecosystem, with high (but not perfect) efficiency.

\section{Conclusions}
\label{sect6}
We have shown how fermionic raising and lowering operators can be  used in the description of a closed ecosystem, in which organisms
living in two (or more) different levels grow up feeded by some nutrients produced by the {\em decay} of some kind of garbages which, in turn, is
produced by the metabolic waste and/or the death of the organisms themselves. The dynamics of the various compartments of the system has been deduced, analytically
or numerically, for the linear models, and numerically for the nonlinear  model.

We have also discussed how our models can be made {\em efficiently decaying}. In particular, we have shown that replacing a real inertia with a
complex parameter, with negative imaginary part, and assuming that the Heisenberg equations of motion still produce the time evolution of the system, we get such a damping.
This has been checked both for the linear and for the nonlinear models.

As already mentioned, ours should be considered as preliminary results, useful to understand whether raising and lowering operators can be of
some utility in the description of biological systems. In our opinion, the results discussed here prove indeed that this is the case. More
sophisticated models will be analyzed soon.

\section*{Acknowledgments}

This work has been financially supported in part by G.N.F.M. of I.N.d.A.M., and by local Research Projects of the Universities of Messina and
Palermo.

\appendix

\renewcommand{\theequation}{A.\arabic{equation}}

\section{Few results on the number representation}
\label{AppendixA}

Let $\Hil$ be an Hilbert space, and $B(\Hil)$ the set of all the bounded operators on $\Hil$.    Let $\ST$ be our physical system, and $\A$ the
set of all the operators useful for a complete description of $\ST$, which includes the \emph{observables} of $\ST$. For simplicity, it is
convenient to assume that  $\A$ coincides with $B(\Hil)$ itself. The description of the time evolution of $\ST$ is related to a self--adjoint
operator $H=H^\dagger$ which is called the \emph{Hamiltonian} of $\ST$, and which in standard quantum mechanics represents  the energy of
$\ST$. We will adopt here the so--called \emph{Heisenberg} representation, in which the time evolution of an observable $X\in\A$ is given by
\be X(t)=\exp(iHt)X\exp(-iHt), \label{a1} \en or, equivalently, by the solution of the differential equation \be
\frac{dX(t)}{dt}=i\exp(iHt)[H,X]\exp(-iHt)=i[H,X(t)],\label{a2} \en where $[A,B]:=AB-BA$ is the \emph{commutator} between $A$ and $B$. The time
evolution defined in this way is a one--parameter group of automorphisms of $\A$.

An operator $Z\in\A$ is a \emph{constant of motion} if it commutes with $H$. Indeed, in this case, equation (\ref{a2}) implies that $\dot
Z(t)=0$, so that $Z(t)=Z$ for all $t$.

In some older papers, see Ref.~\onlinecite{bagbook} and references therein, a special role was played by the so--called \emph{canonical commutation
relations}. Here, as in Ref.~\onlinecite{ff3}, these are replaced by the so--called \emph{canonical anti--commutation relations} (CAR): we say that a set
of operators $\{a_\ell,\,a_\ell^\dagger, \ell=1,2,\ldots,L\}$ satisfy the CAR if the conditions \be \{a_\ell,a_n^\dagger\}=\delta_{\ell
n}\Id,\hspace{8mm} \{a_\ell,a_n\}=\{a_\ell^\dagger,a_n^\dagger\}=0 \label{a3} \en hold true for all $\ell,n=1,2,\ldots,L$. Here, $\Id$ is the
identity operator and $\{x,y\}:=xy+yx$ is the {\em anticommutator} of $x$ and $y$. These operators, which are widely analyzed in any textbook
about quantum mechanics (see,  for instance,
Refs.~\onlinecite{rom,mer}) are those which are used to describe $L$ different \emph{modes} of fermions. From
these operators we can construct $\hat n_\ell=a_\ell^\dagger a_\ell$ and $\hat N=\sum_{\ell=1}^L \hat n_\ell$, which are both self--adjoint. In
particular, $\hat n_\ell$ is the \emph{number operator} for the $\ell$--th mode, while $\hat N$ is the \emph{number operator of $\ST$}.
Compared with bosonic operators, the operators introduced here satisfy a very important feature: if we try to square them (or to rise to higher
powers), we simply get zero: for instance, from (\ref{a3}), we have $a_{\ell}^2=0$. This is related to the fact that fermions satisfy the Fermi
exclusion principle \cite{rom}.

The Hilbert space of our system is constructed as follows: we introduce the \emph{vacuum} of the theory, that is a vector $\varphi_{\bf 0}$
which is annihilated by all the operators $a_\ell$: $a_\ell\varphi_{\bf 0}=0$ for all $\ell=1,2,\ldots,L$. Then we act on $\varphi_{\bf 0}$
with the  operators $a_\ell^\dagger$ (but not with higher powers, since these powers are simply zero!): \be
\varphi_{n_1,n_2,\ldots,n_L}:=(a_1^\dagger)^{n_1}(a_2^\dagger)^{n_2}\cdots (a_L^\dagger)^{n_L}\varphi_{\bf 0}, \label{a4} \en $n_\ell=0,1$ for
all $\ell$. These vectors give an orthonormal set and are eigenstates of both $\hat n_\ell$ and $\hat N$: $\hat
n_\ell\varphi_{n_1,n_2,\ldots,n_L}=n_\ell\varphi_{n_1,n_2,\ldots,n_L}$ and $\hat N\varphi_{n_1,n_2,\ldots,n_L}=N\varphi_{n_1,n_2,\ldots,n_L}$,
where $N=\sum_{\ell=1}^Ln_\ell$. Moreover, using the  CAR, we deduce that $\hat
n_\ell\left(a_\ell\varphi_{n_1,n_2,\ldots,n_L}\right)=(n_\ell-1)(a_\ell\varphi_{n_1,n_2,\ldots,n_L})$ and $\hat
n_\ell\left(a_\ell^\dagger\varphi_{n_1,n_2,\ldots,n_L}\right)=(n_\ell+1)(a_l^\dagger\varphi_{n_1,n_2,\ldots,n_L})$, for all $\ell$. The
interpretation does not differ from that for bosons \cite{bagbook} and then $a_\ell$ and $a_\ell^\dagger$ are again called the
\emph{annihilation} and the \emph{creation} operators. However, in some sense, $a_\ell^\dagger$ is {\bf also} an annihilation operator since,
acting on a state with $n_\ell=1$, we destroy that state.

The Hilbert space $\Hil$ is obtained by taking  the linear span of all these vectors. Of course, $\Hil$ has a finite dimension. In particular,
for just one mode of fermions, $dim(\Hil)=2$. This also implies that, contrarily to what happens for bosons, the fermionic operators are
bounded.

The vector $\varphi_{n_1,n_2,\ldots,n_L}$ in (\ref{a4}) defines a \emph{vector (or number) state } over the algebra $\A$  as \be
\omega_{n_1,n_2,\ldots,n_L}(X)= \langle\varphi_{n_1,n_2,\ldots,n_L},X\varphi_{n_1,n_2,\ldots,n_L}\rangle, \label{a5} \en where
$\langle\,,\,\rangle$ is the scalar product in  $\Hil$. As we have discussed in Ref.~\onlinecite{bagbook}, these states are used to \emph{project} from
quantum to classical dynamics and to fix the initial conditions of the considered system.

\renewcommand{\theequation}{B.\arabic{equation}}

\section{Phenomenological damping}
\label{AppendixB}
The problem of a simple description of irreversible processes in quantum mechanics is usually very hard. Probably, the simplest choice consists
in using a non self-adjoint, effective, hamiltonian which is properly chosen in order to describe the phenomenon we are interested to. For
instance, in Refs.~\onlinecite{benaryeh,tri}, a non self-adjoint two-by-two matrix hamiltonian is used to describe some kind of interactions of a two-level atom
with the radiation. Of course, using such an operator to describe the time evolution of a system usually causes several problems. First of all,
it is not evident at all that the dynamics is still driven by a Heisenberg-like equation of motion. Actually,
in Refs.~\onlinecite{benaryeh,tri}, as well
as in Refs.~\onlinecite{graefe,bagpf2d}, the assumption is that the wave-function $\Psi(t)$ of the system still evolves obeying the Schr\"odinger equation
$i\dot \Psi=H\Psi$, even is $H\neq H^\dagger$. Here, it is more convenient to adopt the dual point of view: the time evolution of the
observable $X$ is still given by equation (\ref{a1}), even if $H\neq H^\dagger$. This choice has consequences on the choice of the {\em natural}
scalar product of the Hilbert space of the theory. These are aspects which we will not consider here, since they are not relevant for us.

In order to find conditions which produce damping we consider the following simple interacting model:
$$
H=\omega_1 a_1^\dagger a_1+\omega_2 a_2^\dagger a_2+\lambda(a_1^\dagger a_2+a_2^\dagger a_1),
$$
where $[a_i,a_j^\dagger]=\delta_{i,j}\,\Id$, $i,j=1,2$, and $\omega_j, \lambda\in\Bbb{R}$, at least for the time being. This model is a linear
version of that introduced in Ref.~\onlinecite{ff2} in connection with love affairs, with an extra term ($H_0=\omega_1 a_1^\dagger a_1+\omega_2
a_2^\dagger a_2$) added to the original hamiltonian, which is useful to introduce the {\em inertia} of the lovers \cite{bagbook}. The time
evolution of $a_j(t)$, and of $\hat n_j(t)=a_j^\dagger(t)a_j(t)$ as a consequence, can be deduced analytically, and the mean values of $\hat
n_j(t)$ can also be found:
$$
\left\{
\begin{aligned}
n_1(t)&=\left<\hat n_1(t)\right>=n_1 |\Phi_{1,1}(t)|^2+n_2 |\Phi_{1,2}(t)|^2,  \\
n_2(t)&=\left<\hat n_2(t)\right>=n_1 |\Phi_{2,1}(t)|^2+n_2 |\Phi_{2,2}(t)|^2.
\end{aligned}
\right.
$$
Here $n_j=\left<\hat n_j(0)\right>$ are fixed by the initial conditions for the system, while the various functions $|\Phi_{k,l}(t)|$ all share
the same general analytic expression:
$$
|\Phi_{k,l}(t)|^2=a_{11}e^{it(\overline{\alpha}_1-\alpha_1)}+a_{12}e^{it(\overline{\alpha}_2-\alpha_2)}+a_{21}e^{it(\overline{\alpha}_2-\alpha_1)}+
a_{22}e^{it(\overline{\alpha}_1-\alpha_2)},
$$
 where $a_{ij}$ are constants, which are not very relevant for us here, while $\alpha_1=\frac{1}{2}\left(\omega_1+\omega_2+\Omega\right)$
 and $\alpha_2=\frac{1}{2}\left(\omega_1+\omega_2-\Omega\right),$ where $\Omega=\sqrt{(\omega_1-\omega_2)^2+4\lambda^2}$. It is clear that,
 as far as $\omega_j$ and $\lambda$ are real, $n_1(t)$ and $n_2(t)$ can only oscillate. On the other hand, let us consider the possibility
 of having these parameters complex-valued: $\omega_j=\omega_{j,re}+i\omega_{j,im}$, $\lambda=\lambda_{re}+i\lambda_{im}$, with $\omega_{j,re}$,
 $\omega_{j,im}$, $\lambda_{re}$ and $\lambda_{im}$ real. A simple analysis suggests that, in order to get $n_j(t)\rightarrow0$
 for $t\rightarrow\infty$, it is enough to add a negative imaginary part to $\omega_1$ or to $\omega_2$. More precisely, if we
 take $\lambda_{im}=0$ and $\omega_{1,im}=\omega_{2,im}<0$, both $n_1(t)$ and $n_2(t)$ goes to zero asymptotically. On the other hand,
 it is easy to check that, if $\omega_{1,im}=\omega_{2,im}=0$, there is no possible choice of $\lambda_{re}$ and $\lambda_{im}$ which
 produces damping.

 This simple model suggests that, for a phenomenological description of damping, it is sufficient to add a (small) negative imaginary part to the
 parameters of the free hamiltonian, leaving unchanged the (real) interaction parameter. It might be worth noticing that it is not important here the fact
 that we are working with bosons: indeed,  the same conclusions could be deduced also working with fermionic operators. In fact,
 this is exactly what we have deduced in Section \ref{sect5}.

\end{document}